\documentclass[]{emulateapj}
\usepackage{apjfonts}
\usepackage{natbib}
\usepackage{graphicx}
\usepackage{epstopdf}
\usepackage{multirow} % per celle tabelle che occupano piu' righe
\usepackage{longtable,lscape} % per tabelle lunghe e in landscape
\usepackage{color}
\usepackage{xspace}
\usepackage{xargs}
\usepackage{url}

\newcommand{\myemail}{benedetta.vulcani@ipmu.jp}

\newcommand{\umgs}{UMGs\xspace}
\newcommand{\umg}{UMG\xspace}
\newcommand{\pr}{progenitor\xspace}
\newcommand{\prs}{progenitors\xspace}
\newcommand{\dl}{DLB07\xspace}

\shorttitle{Progenitors' environment and mass growth}
\shortauthors{Vulcani et al.}

\begin{document}
\title{Mergers and Star Formation: The environment and Stellar Mass Growth of the Progenitors of Ultra-Massive Galaxies since z = 2} %\thanks{}}

\author{Benedetta Vulcani\altaffilmark{1}}
\author{Danilo Marchesini\altaffilmark{2}}
\author{Gabriella De Lucia\altaffilmark{3}}
\author{Adam Muzzin\altaffilmark{4}}
\author{Mauro Stefanon\altaffilmark{5}}
\author{Gabriel B. Brammer\altaffilmark{6}}
\author{Ivo Labb\'e\altaffilmark{5}}
\author{Olivier Le F\`evre\altaffilmark{7}}
\author{Bo Milvang-Jensen\altaffilmark{8}}

\affil{\altaffilmark{1}Kavli Institute for the Physics and Mathematics of the Universe (WPI), The University of Tokyo Institutes for Advanced Study (UTIAS), the University of Tokyo, Kashiwa, 277-8582, Japan; \myemail}
\affil{\altaffilmark{2}Department of Physics and Astronomy, Tufts University, Medford, MA 02155, USA}
\affil{\altaffilmark{3}INAF- Astronomical Observatory of Trieste, I-34143 Trieste, Italy}
\affil{\altaffilmark{4}Institute of Astronomy, University of Cambridge, Madingley Road, Cambridge CB3 0HA}
\affil{\altaffilmark{5}Leiden Observatory, Leiden University, P.O. Box 9513, 2300 RA Leiden, The Netherlands}
\affil{\altaffilmark{6}Space Telescope Science Institute, 3700 San Martin Drive, Baltimore, MD, 21218, USA}
\affil{\altaffilmark{7}Aix Marseille Universit\'e, CNRS, Laboratoire d'Astrophysique de Marseille, UMR 7326, 13388, Marseille, France}
\affil{\altaffilmark{8}Dark Cosmology Centre, Niels Bohr Institute, University of Copenhagen, Juliane Maries Vej 30, DK-2100 Copenhagen, Denmark}

\begin{abstract}
The growth of galaxies is a key problem in understanding the structure and evolution of the universe. 
Galaxies grow their stellar mass by a combination of 
star formation and mergers, with a relative importance that is redshift dependent.  Theoretical models predict quantitatively different contributions from the two channels;  measuring these from the data is a crucial constraint.
Exploiting the UltraVISTA catalog and a unique sample of \prs of local ultra massive galaxies selected with an abundance matching approach, we quantify the role of the two mechanisms from $z=2$ to 0.  We also compare our results to two independent incarnations of semi-analytic models.
At all redshifts, progenitors are found in a variety of environments, ranging from being isolated to having 5-10 companions with mass ratio at least 1:10 within a projected radius of 500 kpc. In models, \prs have a systematically larger number of companions,  entailing a larger mass growth for mergers than in observations, at all redshifts.
 Generally,   in both observations and  models, %, small discrepancies are found between 
the inferred  and the expected mass growth roughly agree, within the uncertainties.
Overall, our analysis confirms the model predictions, showing how the growth history of massive galaxies is dominated by in situ star formation at $z\sim2$, both star-formation and mergers at $1<z<2$, and by mergers alone at $z<1$.
Nonetheless, detailed comparisons still point out to tensions between the expected mass growth 
and our results, 
which might be due to either an incorrect progenitors-descendants selection, uncertainties on star formation rate and mass estimates, or the adopted assumptions on  merger rates.

\end{abstract}

\section{Introduction}
Even though in the last decades much attention has been dedicated to the study of galaxy formation and evolution, understanding when and how the most massive galaxies formed and how they evolve with time are still  controversial questions. In the standard paradigm of structure formation, dark matter halos assemble hierarchically in a gravitational collapse, and galaxies form inside these structures following the radiative cooling of baryons.  Stars in today's most massive galaxies ($M_\ast\sim 10^{12} M_\odot$) are formed very early (50\% at $z\sim$5, 80\% at $z\sim$3) and in many small galaxies. Model massive galaxies  can have a number of effective progenitors as high as $\sim$5 and assemble surprisingly late. Predictions are model dependent; e.g. according to \cite{delucia06, deluciablaiz07} half their final mass is typically locked-up in a single galaxy after $z\sim$0.5.

Many physical processes have to be taken into account to explain the growth of massive galaxies. Star formation is  expected to play an important role at higher redshifts as a large fraction of massive galaxies at $z\sim 2$ have high star formation rates \citep[e.g,][]{vd04, papovich06}. However, the old stellar ages of the most massive early-type galaxies \citep[e.g,][]{thomas05, vd07} and the existence of apparently ``red and dead'' galaxies with small sizes at $z = 1.5-2.5$ \citep[e.g,][]{cimatti08, vd08} suggest that at least some of the growth is due to other mechanisms, like mergers.
Recently, \cite{graham13, dullo13, graham15} have also suggested that some massive galaxies have evolved by accreting a large disc of gas which rapidly forms stars rather than growth only via mergers.  

Below $z\sim 1$, most massive galaxies ($M_\ast\sim 10^{12} M_\odot$) are generally  found in dense environments \citep{blanton09}, such as at the center of clusters and are identified as the Brightest Cluster Galaxies (BCGs).  Nonetheless, the recent survey MASSIVE \citep{ma14} showed that this might not be always the case. They observed 116 galaxies with $M_\ast\sim 10^{11.5} M_\odot$  and distance $D < 108 Mpc$ ($z < 0.025$) finding that  48-56\% of them are located in groups (35-38\% are actually central galaxies), while 6-14\% are isolated. 
Many studies  focused on characterizing the assembly of massive galaxies, both from a theoretical and observational point of view. \cite{lidman12, lin13} found that the stellar mass of BCGs increases by a factor of $\sim$ 2 since z$\sim1$. Star formation rates in BCGs at z$\sim$1 are generally too low to result in significant amounts of mass. Instead, most of the mass build up occurs through  mergers.
In semi-analytic models, accretion of satellite galaxies are mainly dry and minor \citep[e.g.,][]{deluciablaiz07}, while in observations many works point to  major mergers in the center of clusters as the main responsible for the mass growth \citep{rasmussen10, brough11, bildfell12, lidman13}.  Characterizing  three BCGs at z$\sim$0.1 with nearby companions, \cite{brough11} found that the companions of two of the BCGs would merge with the BCG within 0.35 Gyr. Additional examples of likely major mergers can be found in \cite{rasmussen10, yamada02, collins09}. While it is clear that mergers do occur, it is not yet clear what fraction of the stars in the merging galaxies ends up in the BCG and what fraction is distributed throughout the cluster.  High resolution simulations suggest that between 50 to 80\% of the mass of mergers is not locked into  galaxies, but is  distributed throughout the cluster \citep{conroy07, puchwein10}.
Recent observational studies, on the basis of color gradients, support simulations, claiming  that at least half of the mass is lost into the intracluster medium of the clusters \citep[e.g.,][]{lidman12, burke15}.

At higher redshift, the environment in which massive galaxies reside is less characterized, and a clear correspondence between massive galaxies and BCGs is lacking. Nonetheless, a number of studies characterized  the build up of massive galaxies.  \cite{ownsworth14}, using a variety of number density selections, claimed that more than half of the total stellar mass in massive galaxies ($M_\ast \sim 10^{11.24}M_\odot$) at $z = 0.3$ is created externally to their $z = 3$ progenitors. \cite{vdokkum10}, selecting galaxies at a constant number density of $n=2 \times  10^{-4} Mpc^{-3}$, found that the role of mergers might be even more important, with star formation accounting only for 20\% of the total mass growth. 
Connecting high and low redshift BCG data via evolution of their host halo masses, \cite{shankar15} found an increase since $z\sim 1$ of a factor $\sim$ 2-3 in their mean stellar mass and $\sim$ 2.5-4 in their mean effective radius.

To really understand how individual galaxies have evolved and assembled their mass and what mechanisms drive these changes, it is important to  properly connect today's most massive galaxies to their \prs at earlier times. This  requires the non-trivial task of linking galaxies and their descendants/progenitors through cosmic time, which in turn requires assumptions for how galaxies evolve. In recent years, a few approaches have been developed to link galaxies across cosmic time \citep[e.g.,][]{brammer11, conroy09, behroozi13, wake06, leja13, mundy15}. Whereas the limitations of these techniques are still being debated \citep[e.g.,][]{torrey15}, it is widely recognized that these approaches are far superior than selecting galaxies at fixed stellar mass for studies of galaxy evolution. 

\cite{marchesini14} assembled the first sample of galaxies defined to be the \prs of galaxies with $\log M_\ast/M_\odot > 11.8$ at $z=0$ (ultra-massive galaxies, hereafter \umgs)
from $z = 3$ using a semi-empirical approach based on abundance matching in the $\Lambda$CDM paradigm \citep[see \S \ref{sec:selection}]{behroozi13}. Characterizing the stellar population properties of the \prs (masses, ages, dust star formation), they claimed that at least half of the stellar content of local \umgs was assembled at $z > 1$, whereas the remaining was assembled via merging from $z \sim 1$ to the present. They also found that most of the quenching of the star-forming progenitors happened between $z = 2.75$ and $z = 1.25$, in good agreement with the typical formation redshift and scatter in age of $z = 0$ \umgs as derived from their fossil records. The progenitors of local \umgs, including the star-forming ones, never lived on the blue cloud since $z = 3$.

Using the unique  progenitor-descendant sample presented in \cite{marchesini14}, in this paper we focus on the environment in which these \prs reside, and test whether their mass growth can  be ascribable mainly to mergers or to star formation.
In particular, we explicitly test the model predictions for the different contributions to the stellar mass assembly since $z\sim$2. Both semi-analytic models \citep[e.g.,][]{zehavi12} and abundance matching techniques based on halo occupation models \citep[e.g.,][]{conroy09} indicate that star formation is important at all halo masses at z$\sim2$, at $z<1$ accretion through mergers dominates at the high-mass end ($\sim 10^{13} h^{-1} M_\odot$) of the halo mass distribution, where star formation is negligible, while at intermediate redshift both contributions are important.  Solving possible discrepancies found between observations and simulations is beyond the scope of this paper, and is deferred to a forthcoming analysis. 

We parameterize the environment in terms of projected distance from the \pr, since we have no information about the mass of the haloes these galaxies reside.  We only consider  mergers  between galaxies with a mass ratio at most of 1:10. We  also compare our observational results to the predictions of two semi-analytic models, namely the \citet[hereafter \dl]{deluciablaiz07} and the  \citet[hereafter H15]{henriques15} model, to investigate whether the \cite{marchesini14} approach to link galaxies across the cosmic time is supported by these models.

Throughout the paper, we assume $H_{0}=70 \, \rm km \, s^{-1} \,
Mpc^{-1}$, $\Omega_{0}=0.3$, and $\Omega_{\Lambda} =0.7$.  We adopt a  \cite{kroupa01}
initial mass function (IMF) in the mass range 0.1--100
$M_{\odot}$. 

\section{Data set}\label{sec:sample}
Our sample is drawn from  the $K_S$-selected catalog of the COSMOS/UltraVISTA field from \citeauthor{muzzin13} (2013a). The catalog  covers 1.62 $deg^2$ and includes point-spread function-matched photometry in 30 photometric bands over the wavelength range 0.15$\mu m$-24$\mu m$ from the available GALEX \citep{martin05}, Canada-France-Hawaii Telescope/Subaru \citep{capak07}, UltraVISTA \citep{mccracken12}, and S-COSMOS \citep{sanders07} data sets. Sources are selected from the DR1 UltraVISTA $K_S$-band imaging \citep{mccracken12} which reaches a depth of $K_{S,tot} <$ 23.4 at 90\% completeness. Details on the photometric catalog construction, photometric redshift measurements, and stellar population properties' estimates can be found in \citeauthor{muzzin13} (2013a).  
Briefly, stellar population properties were derived by fitting the observed spectral energy distributions (SEDs) from the GALEX UV to the Spitzer-IRAC 8$\mu m$ photometry with \cite{BC03} models assuming exponentially declining SFHs of the form $SFR\propto e^{-t/\tau}$, where $SFR$ is the Star Formation Rate, $t$ is the time since the onset of star formation, and $\tau$ sets the timescale of the decline in the SFR, solar metallicity, a \cite{calzetti00} dust law, and a \cite{kroupa01} IMF \citep[see also][]{marchesini14}. 

 \cite{marchesini14} investigated the effects of different SED-modeling assumptions by adopting, among others,  different SFHs, and metallicities. Adopting a delayed-$\tau$ SFH  in place of the exponentially-declining SFH allows for increasing SFR at earlier times. The delayed-$\tau$ model implies SFRs smaller by $\sim$0.1 dex and  stellar ages $\langle t \rangle_{SFR}$ larger by $\tau \sim$0.1 dex compared to the default SED- modeling assumptions. Relaxing the assumption on metallicity by leaving it as a free parameter in the SED modeling does not noticeably change the result, indicating that the impact of fixing the metallicity to the solar one is almost negligible. 
Overall, the systematic effects on the stellar population properties are found to be significantly smaller than the corresponding typical random uncertainties for most of the different SED-modeling assumptions.  Therefore, results are robust and not very sensitive to reasonable choices of the SED-modeling assumptions.

The redshift-dependent stellar mass completeness limit has been presented in  \citeauthor{muzzin13} (2013b). This was determined by selecting galaxies belonging to the available deeper samples and then scaled  fluxes and $M_\star$ to match the K-band completeness limit of the UltraVISTA sample ($K_{S,tot}=$ 23.4). The upper envelope of points in the ($M_{\star,scaled} - z$) space represents the most massive galaxies at $K_S$ = 23.4, and so provides a redshift-dependent 100\% $M_\star$ completeness limit for the UltraVISTA sample. Similarly, \citeauthor{muzzin13} (2013b) also derived 95\% mass-completeness limits for the sample, which increases the sample by a factor of 1.4. Given this substantial increase in statistics, we follow \citeauthor{muzzin13} (2013b) and adopt the 95\% mass-completeness limits.

The quiescent/star- forming separation has been done using the rest-frame U-V versus V-J color-color diagram and is presented in \citeauthor{muzzin13} (2013a), \cite{marchesini14}. This method has the ability to separate red galaxies that are quiescent from reddened (i.e., dust-obscured) star-forming galaxies (see, e.g., \citealt{labbe06, wuyts07, williams09, brammer11, patel11, whitaker11}; \citeauthor{muzzin13} 2013a). 

\subsection{The selection of the progenitors of local ultra-massive galaxies}\label{sec:selection}
The selection of the progenitors of local \umgs has been performed adopting a semi-empirical approach which uses abundance matching in the $\Lambda$CDM paradigm \citep[see][]{marchesini14}. This method  accounts for mergers and scatter in mass accretion histories. Details on this technique can be found in \citet[and references therein]{behroozi13}. Briefly, the galaxy cumulative number density at redshift $z_1$ is converted to a halo mass with equal cumulative number density using peak halo mass functions. Then, for halos at that mass at $z_1$, the masses of the most-massive progenitor halos at $z_2 > z_1$ are recorded using to the halos' mass accretion histories. Finally, the median halo progenitor mass at $z_2$ is converted back into cumulative number densities using the halo mass function at $z_2$. %  \citep{marchesini14}.

The progenitors of the low-$z$ population of very massive galaxies are traced by identifying, at each redshift, the stellar mass for which the evolving cumulative number density intersects the cumulative number density curves derived from the UltraVISTA stellar mass functions (\citeauthor{muzzin13} 2013a). 
In this way, a sample of progenitors of galaxies with a mass of $\sim 10^{11.8} M_\odot$ at $z\sim0$ is assembled.

 The typical error on the progenitors' stellar mass resulting from the uncertainties of the observed stellar mass functions and cumulative number densities was found to be in the range 0.03-0.07 dex \citep{marchesini14}. The inferred growth in stellar mass of the progenitors was therefore found to be 0.45$\pm$0.13 dex and 0.27$\pm$0.08 dex from $z=2$ and z=1, respectively, to $z=0$. If the scatter in mass accretion histories is also included in the error analysis, the uncertainties on the inferred growth in stellar mass of the progenitors increase by a factor of $\sim$1.7 \citep{marchesini14}.

\subsection{Our sample}
\begin{figure}
\centering
\includegraphics[scale=0.4]{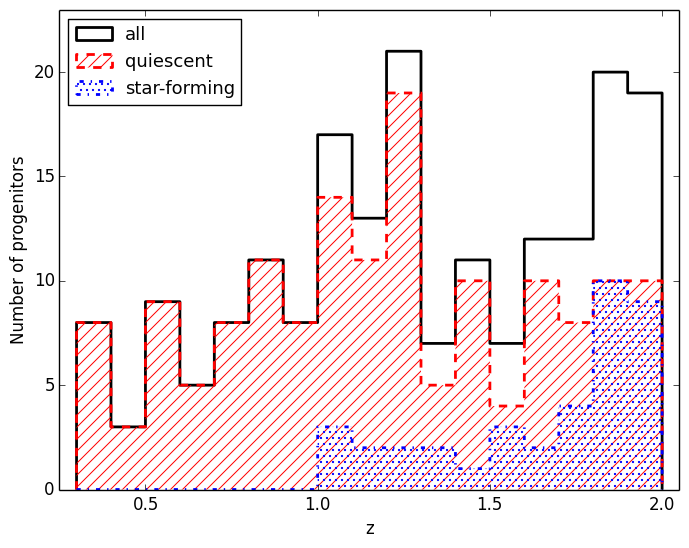}
\caption{Redshift histogram of our \prs of \umgs sample (black). The distribution of quiescent (red) and star-forming (blue) galaxies is also shown.  \label{distr_prog}}
\end{figure}
In this work, we use the sample of \prs of \umgs defined in \cite{marchesini14}, but we limit our analysis to galaxies at $z<$2. In this way, at all redshifts, the UltraVISTA sample includes all galaxies at least as massive as 1:10 the mass of the closest \pr in the redshift range $\pm 0.05\times(1+z_{pr})$ with $z_{pr}$ the photo-z of the \pr. The  redshift range is chosen to take into account the typical photo-$z$ accuracy, which is redshift dependent \citep[see, e.g., ][]{muzzin13b}.  191 \prs enter our selection. As shown in Fig.~\ref{distr_prog}, the number of \prs depends on redshift: there are 11 galaxies at $0.2<z<0.5$, 41 at $0.5<z<1$, 69  at $1<z<1.5$ and 70  at $1.5<z<2$. The different frequency is mainly due to the different volume covered by the different redshift bins, which is $\sim 10\times$ larger at $z\sim1.75$ than at $z\sim 0.35$. 
 Indeed, the volume probed is $\sim$25 Gpc$^3$ at $0.2<z<0.5$, $\sim$125 Gpc$^3$ at $0.5<z<1$, $\sim$200 Gpc$^3$ at $1<z<1.5$, and $\sim$235 Gpc$^3$ at $1.5<z<2$.

As already pointed out by \cite{marchesini14}, for $z<1$ all galaxies are quiescent, while at higher redshift the fraction of star-forming \prs is not negligible. 

To characterize the environment in which \prs are embedded,  we also make use of the entire UltraVISTA catalog, to which we apply a mass cut to ensure 95\% of completeness (see \S\ref{sec:sample}). 

Hereafter, we will refer to those galaxies around the \prs within a specified projected radius, at least as massive as 1:10 the mass of the \pr and in the redshift range $\pm 0.05\times(1+z_{pr})$ as {\it companion} galaxies. 

We note that there might be systematic effects in the data that alter the robustness of the results, such as  systematic errors in photometric redshifts and contamination of the photometry from emission lines. The latter might result in  overestimates of stellar masses. 
\cite{marchesini14} already investigated
the possible systematics in the UltraVISTA sample, showing that at the redshifts here considered they should not impact on our findings (see also \citeauthor{muzzin13} 2013a).

\section{Theoretical predictions}

  In this paper we compare the findings from the observations with the  predictions from theoretical models on the environment of the \prs  of today's \umgs and on the relative importance of merging vs. in-situ star formation to their inferred growth  in stellar mass to investigate the physical processes implemented in the models. To this aim, the model predictions are derived in two different ways. In one case, the model-predicted assembly histories of the progenitors are directly exploited to determine the overall growth in stellar mass of the descendants. In the other case, the growth in stellar mass is obtained following the same assumptions used in the observations.

We exploit galaxy catalogs from two semi-analytic models run on the Millennium Simulation \citep{springel05}. This uses  $10^{10}$ particles of mass
$8.6 \times  10^8 \, h^{-1} \, M_\odot$ to trace the evolution of the matter distribution in a cubic region of the Universe of $500\, h^{-1} \, Mpc$ on a side from $z = 127$ until $z = 0$, and has a spatial resolution of $5\, h^{-1} \, kpc$.  

We use two different semi-analytic models to investigate
how different assumptions about 
the physical processes acting on the baryonic component
impact the evolution of the galaxy  masses.

The semi-analytic model discussed in 
\dl builds on the methodology and prescriptions introduced in \cite{springel01, delucia04, croton06} and has been the first variant of the ``Munich'' models family that has been made publicly available.  
The \dl model is based on WMAP1 cosmology \citep{sanchez06} and includes prescriptions for supernova-driven winds, follows the growth of supermassive black holes and includes a phenomenological description of AGN feedback. 
The model neglects environmental physical processes such as ram pressure and harassment, but assumes that when galaxies are accreted on to a more massive system, the associated hot gas reservoir is stripped instantaneously. This induces a very rapid decline of the star formation histories of satellite galaxies, and contributes to create an excess of red and passive galaxies with respect to the observations \citep[e.g.,][]{wang07}.
The \dl model is mainly tuned to reproduce  the K-band luminosity function at z=0.

We also use the model presented in H15, which represents one of the most recent updates of the Munich models.
The H15 model uses the Planck first-year cosmology and
basically contains the same physics as the \dl model but has a more sophisticated treatment for the evolution of satellites. 
Differently from the \dl model, which does not include a channel for ICL formation, the H15 model includes tidal stripping as a channel for ICL. In addition, it adds a modification of the time-scale to re-accrete gas ejected through galactic winds and modifies the  ram-pressure stripping in halos less massive than $\sim10^{14} M_\odot$.
The model is tuned to reproduce recent data on the abundance and passive fractions of galaxies and the galaxy stellar mass function from $z = 3$ down to $z = 0$.
We refer to the original papers for more details.

As explained in \cite{springel01, delucia04a}, models make a distinction between centrals, satellites and orphans. Centrals (type 0) are located at the position of the most bound particle in their halo. These galaxies are fed by gas cooling from the surrounding hot halo medium. Satellites (type 1) were previously central galaxies of another halo, which then merged to form  the larger system in which they currently reside. For these galaxies gas is no longer able to cool on to halo galaxies.
Orphans (type 2) are galaxies no longer associated with distinct dark matter substructures, and in the \dl model their stellar mass is  not affected by the tidal stripping that reduces the mass of their parent halos. In the H15 model such orphans are unable to retain gas ejected by supernova feedback, which is moved to the hot halo of the galaxy group. Tidal forces can completely disrupt the stellar and cold gas components of orphan galaxies, which are then added to the intra-cluster light and the hot gas atmosphere of the group/cluster, respectively.
In both models, orphans may later merge into the central galaxy of their halo. In our analysis, when useful,  we will  distinguish among the three types of galaxies. 

\subsection{Our sample}
\begin{table*}
\caption{Percentages of type 0, type 1 and type 2 galaxies among the progenitors and descendants in the two semi-analytic models. }
\begin{center}
\begin{tabular}{c|cccccc|cccccc}
 & \multicolumn{6}{c|}{{\bf DLB07}}															& \multicolumn{6}{c}{{\bf H15}}	 \\
{\bf z} & \multicolumn{3}{c}{{\bf \% progenitors}}	& \multicolumn{3}{c|}{{\bf \% descendants}}	& \multicolumn{3}{c}{{\bf \% progenitors}}	& \multicolumn{3}{c}{{\bf \% descendants}} \\
\hline
		& type 0 	& type 1 & type 2 	&  type 0 	& type 1 & type 2&  type 0 	& type 1 & type 2	& type 0 	& type 1 & type 2 \\
0.36	& 91$^{+1}_{-1}$		& 8$^{+1}_{-1}$			& 1.2$^{-0.4}_{+20.6}$		& 88$^{+1}_{-1}$ 		& 6$^{+1}_{-1}$			& 6$^{+1}_{-1}$ 		& 85$^{+1}_{-1}$		& 12$^{+2}_{-1}$			& 2.9$^{+0.8}_{-0.7}$		& 85$^{+2}_{-2}$		& 11$^{+1}_{-1}$			& 4.1$^{+1.0}_{-0.8}$	\\
0.76 	& 94$^{+1}_{-1}$		& 6$^{+1}_{-1}$			& 0.5$^{+0.7}_{-0.3}$		& 87$^{+2}_{-2}$		& 6$^{+1}_{-1}$			& 7$^{+1}_{-1}$ 		& 87$^{+1}_{-1}$		& 10$^{+1}_{-1}$			& 2.9$^{+0.8}_{-0.7}$		& 86$^{+1}_{-1}$		& 10$^{+1}_{-1}$			& 4.3$^{+0.9}_{-0.8}$ \\
1.28	& 93$^{+2}_{-2}$		& 6$^{+2}_{-1}$			& 0.9$^{+1.0}_{-0.5}$		& 85$^{+2}_{-3}$		& 7$^{+2}_{-2}$			& 8$^{+2}_{-2}$ 		& 88$^{+1}_{-1}$		& 8.1$^{+0.9}_{-0.8}$			& 4.2$^{+0.7}_{-0.6}$		& 84$^{+1}_{-1}$		& 12$^{+1}_{-1}$			& 4.5$^{+0.7}_{-0.6}$ \\
1.77	& 95$^{+2}_{-2}$		& 4$^{+2}_{-1}$			& 0.6$^{+1.0}_{-0.5}$		& 85$^{+3}_{-3}$		& 9$^{+3}_{-2}$			& 6$^{+2}_{-2}$ 		& 90$^{+1}_{-1}$		& 6.5$^{+0.9}_{-0.8}$			& 3.1$^{+0.6}_{-0.6}$		& 85$^{+1}_{-1}$		& 11$^{+1}_{-1}$			& 4.6$^{+0.8}_{-0.7}$	\\
\end{tabular}
\end{center}
\label{tab_type}
\end{table*}%
For both the \dl and the H15 models, we extract from the available catalogs all the galaxies at $z=0.36$, 0.76, 1.28 and 1.77 with stellar mass in the same mass range spanned by the \prs at the corresponding redshift (approximately within $\pm 0.15$ the median mass of the \prs). 

Since observed masses might be characterized by systematic errors, from the \dl model we also extracted samples of galaxies to test the impact of these errors. We   assigned to each galaxy mass  in the models a random Gaussian error with width $0.03 \times (1+z)$ \citep[following ][]{ilbert13} and then considered only those galaxies whose perturbed mass is in the mass range  $\pm 0.15$ the stellar mass of the \prs at the corresponding redshift. We performed the entire analysis using both samples, without finding noticeable differences between the results. Therefore, in the following, we will present only the analysis performed on the sample with the original masses from the models.

Overall, in the \dl (H15) model 1027 (1076) galaxies have been extracted at $z=0.36$, 608 (1231) at $z=0.76$, 447 (1969) at $z=1.28$ and 311 (1732) at $z=1.77$.

From the models, we also extract the information regarding the merger trees and the descendants of these galaxies down to $z\sim 0$, with the aim to investigate the real mass growth predicted by the models. 
In addition, we also get the virial mass of the haloes in which these galaxies reside and those of their descendants, to further characterize from a theoretical point of view the \prs' hosting environment. 

Finally, we select all galaxies within a box of 1 physical Mpc on a side, centered on each massive galaxy considered, in order to characterize also the environment of the \prs of \umgs in the same way as in observations.

\section{Descendants of the progenitors in the models}\label{sec:types}
First of all, we can test  whether the approach adopted by \cite{marchesini14} to link galaxies across the cosmic time is supported by the models.

\begin{figure}
\centering
\includegraphics[scale=0.35]{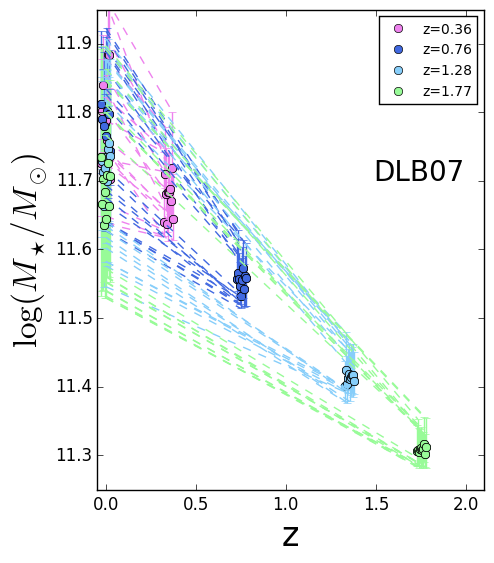}
\includegraphics[scale=0.35]{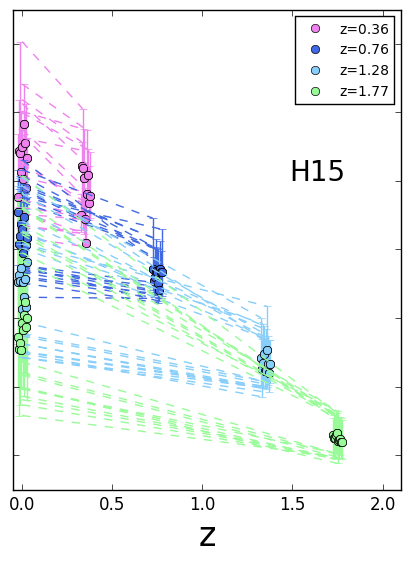}
\caption{Predicted median \pr mass growth by semi analytic models (left: DLB07, right: H15) for galaxies in the same mass range of our \prs. 
Masses at $z=0$ are the masses of the descendants.
Error bars on the y-axis represent  the 25$^{th}$ and 75$^{th}$ percentiles. At each redshift, 10 random extractions have been performed. 
 \label{sim_growth}}
\end{figure}

From simulations, we randomly extract the same number of \prs found in observations at the corresponding redshift and compute the median mass of their descendants at $z\sim0$. We repeat the sampling ten times, to take into account sample variance. Figure \ref{sim_growth} shows that the selection based on the abundance matching method does indeed select galaxies whose mass evolution is consistent with what expected from the \dl model: \prs at the different redshifts will turn into galaxies at z$\sim0$ with masses of $\sim 10^{11.8} M_\odot$, that is the mass inferred by \cite{marchesini14}. The model also reproduces the inferred stellar mass at the intermediate redshifts.  In contrast,  in the H15 model, the median mass of the descendants at $z=0$ is lower than $\sim 10^{11.8} M_\odot$ in most of the extractions, even though within a large dispersion. It is interesting to note that the abundance matching method does not seem to work for the H15 model, which instead is the model that should be in better agreement with the statistics of the halo occupation model by construction (see H15 for details).

Differences between the two models might be due to the fact that the H15 has introduced  tidal stripping and therefore has a smaller number of satellites, producing a smaller mass growth through mergers.

In the following sections we will quantify the separate role of star formation and mergers in the galaxy mass growth, also from a purely theoretical point of view, by explicitly inspecting the merger trees of a subsample of galaxies.

Table \ref{tab_type} shows the percentage of galaxies of a given type, for all objects extracted from the simulations. 
In the \dl model, at all redshifts, the vast majority ($>$90\%) of the \prs are type 0, with the fraction slightly decreasing going from higher to lower redshifts. This finding is probably due to the fact that at lower redshift the number of satellites is larger and that these galaxies  had more time to grow.
It suggests that while at higher redshift massive galaxies are most likely at the center of their halo, in the local universe there is a larger fraction of massive galaxies that are satellites. Only $\leq 1\%$ are type 2. Among the descendants, the fraction of type 0 is smaller and decreases with increasing redshift of the \prs, ranging from 85\% to 88\%. 
In contrast, the fraction of type 2 galaxies  is much larger and they are as common as satellites ($\sim 7\%$ at all redshifts). This suggests that the while almost all massive galaxies that are central at $z=0.36$ will be central also at  $z=0$, the probability of central galaxies to turn into satellites increases with increasing redshift. 

The H15 model presents a systematically lower fraction of type 0 galaxies  (85-90\%) and a slightly larger fraction of type 1 among the \prs, and a similar fraction of type 0 and higher of type 1 among the descendants.

\begin{figure}
\centering
\includegraphics[scale=0.34]{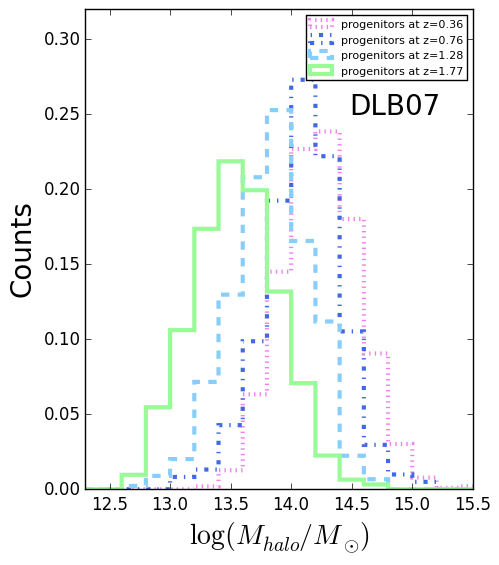}
\includegraphics[scale=0.34]{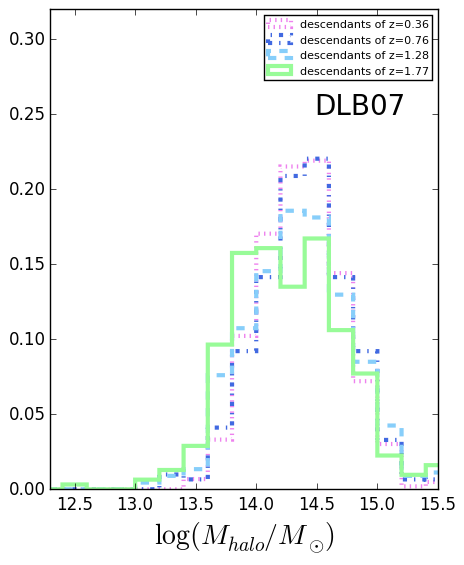}
\includegraphics[scale=0.34]{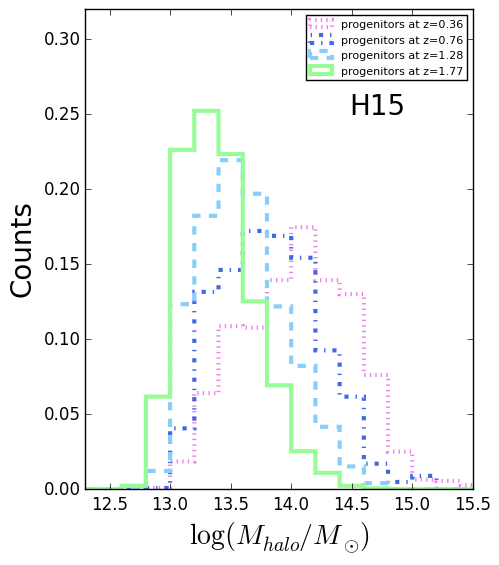}
\includegraphics[scale=0.34]{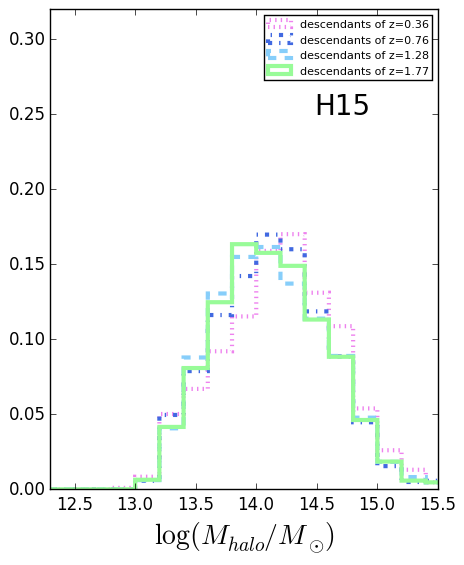}
\caption{Normalized halo mass distribution for \prs at different redshfits (left panels) and for their descendants at $z=0$ (right panels), as indicated in the labels. Upper panels: \dl model, bottom panels: H15 model. \label{halo}}
\end{figure}

Figure \ref{halo} shows the halo mass distribution of all the selected progenitors and also the mass distribution of the haloes where the descendants of the progenitors reside, for both models. 
Both in the \dl and H15 prescriptions, the typical mass of the haloes hosting the \prs clearly increases with decreasing redshifts: galaxies at $z=1.77$ are found in haloes of mass $\sim 10^{12.2}-10^{14.6} M_\odot$ for the \dl model, and of  $\sim 10^{12.7}-10^{14.7} M_\odot$ for the H15 model; galaxies at $z=0.36$ are found in haloes of mass $\sim 10^{13.4}-10^{15.5} M_\odot$ for the \dl model and of $\sim 10^{13}-10^{15.2} M_\odot$ for the H15 model; supporting the finding that the growth of the massive galaxies is coupled to the growth of their haloes \citep[e.g,][and references therein]{tinker12}.
Looking at the halo mass distributions for the descendants,  in the \dl model, they span a similar halo mass range, even though the peaks of the distributions slightly depend on the redshift of the \prs: descendants of the $z$=1.77 \prs are found in slightly less massive haloes than the descendants of the $z$=0.36 \prs (medians values are $M_{vir} = 10^{14.24\pm 0.02}$, and $M_{vir} = 10^{14.37\pm0.01}$, respectively). Therefore, even though the \prs turn into massive galaxies of similar mass, they actually might not end up in the very same galaxies, as expected by the fact that the  growth of structure is stochastic. It also might suggest that also the environment around galaxies should be taken into account when linking galaxies across time, and not only the stellar mass. 
In contrast, in the H15 model all the descendants span a similar halo mass range ($M_{vir} \sim 10^{13}-10^{15.5}$) with a median halo mass of  $10^{14.1\pm 0.01}$. Even though in the H15 model not all the \prs will end up in massive galaxies as selected by the abundance matching technique, these distribution show that they will end up in very similar environments. 

\section{The environment around the progenitors of \umgs}
\begin{figure*}
\centering
\includegraphics[scale=0.32]{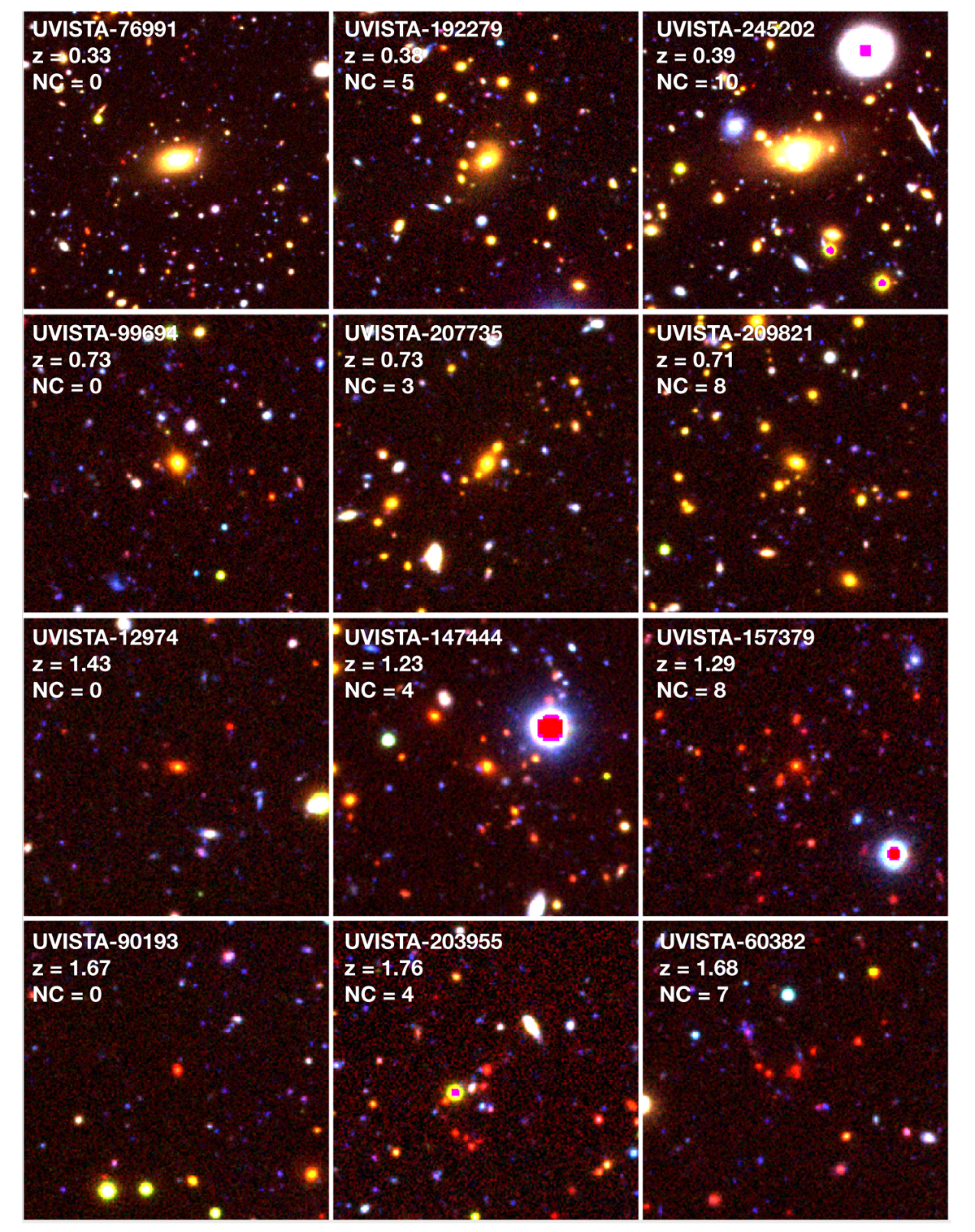}
\caption{Examples of BzK images for isolated galaxies within 250 kpc (first  column), galaxies with few companions (second column) and galaxies with a large number of companions, in four bins of redshifts. Companions are galaxies around the \prs, at least as massive as $\sim$1:10 the mass of the \pr and in the redshift range $\pm 0.05\times(1+z_{pr})$. The UltraVISTA id, the redhisft and the number of companions is indicated in the labels. The FoV of each thumbnail corresponds to a radius of 250 kpc. \label{thumb}}
\end{figure*}
In order to understand the processes that induce the observed galaxy mass growth, we first characterize the environment in which the \prs of \umgs reside. If these galaxies are found in overdense regions, they should easily undergo  mergers; on the other hand, if they are most likely isolated, their growth should be attributable to other factors, such as in-situ star formation. 

Figure \ref{thumb} shows some examples of false color images in the BzK filters covering a FoV of 500 kpc on a side of  galaxies residing in different environments at different redshifts. In order to demonstrate the range of environments of these galaxies, for each redshift bin, we selected a galaxy with no other companions  within a projected sphere of 250 kpc, a galaxy with 3-4 companions a galaxy with $\sim$8-10 companions. Clearly, \prs reside in a variety of environments, which will have a different role in their growth throughout the cosmic time. 

\subsection{The total number of satellites around progenitors}
\begin{figure*}
\centering
\includegraphics[scale=0.55]{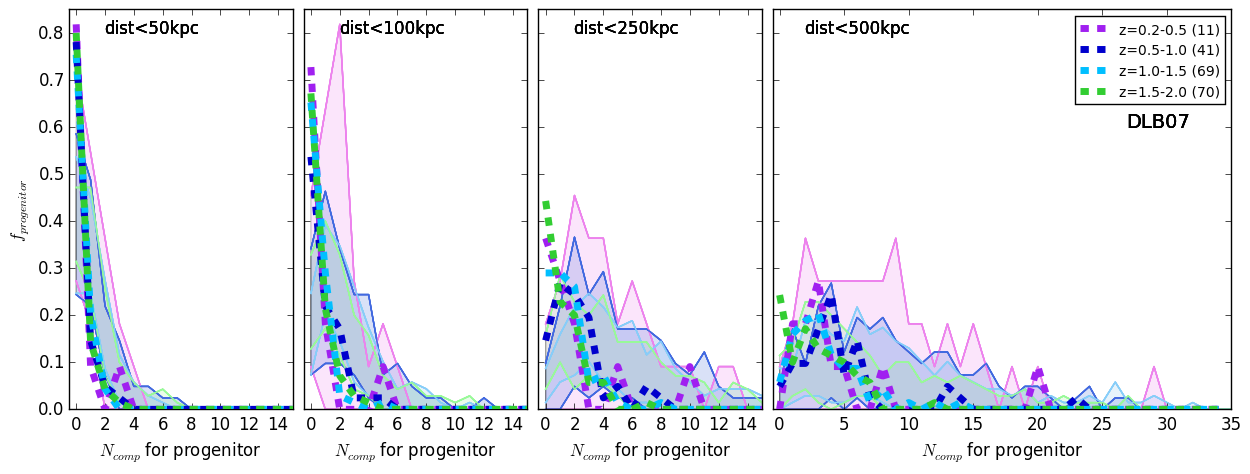}
\includegraphics[scale=0.55]{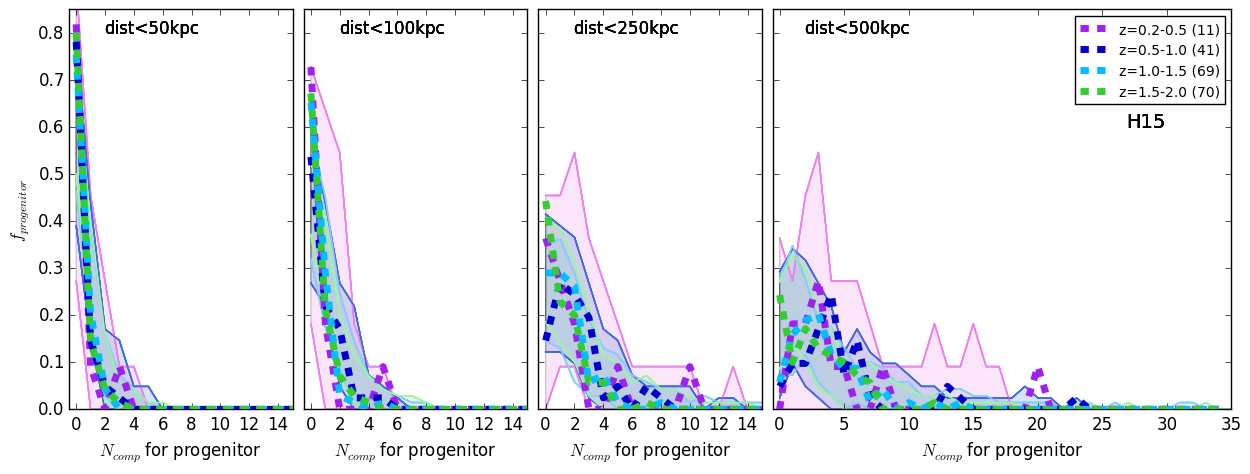}
\caption{Normalized distribution of the number of companions (=galaxies at least as massive as 1:10 the mass of the \pr and with z within $\pm0.05\times(1+z_{pr})$) at different redshifts and within different radii (as indicated on the top of each panel). Thick dashed lines represent observations; thin solid lines and shaded areas represent models (upper panels: DLB07, bottom panels: H15). Numbers in parenthesis give the number of \prs in each redshift bin. \label{distr_sat}}
\end{figure*}

Using the observed sample drawn from the UltraVISTA catalog, we  compute the number of companions around each \pr at different redshifts. 
We consider 
portions of sky centered on the \pr and of different physical radii and   count the number of companions, namely galaxies with mass ratio $>$1:10 and redshift within the range $\pm 0.05\times(1+z_{pr})$, that fall into the projected area. 

Figure \ref{distr_sat} shows the results considering galaxies within 50, 100, 250 and 500 kpc from the \pr, respectively. According to \cite{munoz11},
haloes with $M_{vir}\sim 10^{13} M_\odot$ have typically a  virial radius of $\sim 0.5$ Mpc
below $z\sim 1$, and of $\sim 0.35$ Mpc at $z=2$, while haloes with $M_{vir}\sim 10^{11} M_\odot$ have typically a  virial radius $<0.15$ Mpc at all redshifts. Our binning has been chosen to inspect regions of sky that correspond to the virialized region around the progenitors at the different redshift, assuming they are located in a variety of haloes. The smallest distance indeed should sample the virial radius for any of our progenitor galaxy, while the largest should sample well beyond it.

To get rid of the different volumes covered at the different redshifts and therefore the different number of \prs, for each sample values are normalized to the total number of \prs at the considered redshift. 
In observations (thick lines in both panels), at any redshift, $\sim$80\% of \prs have no galaxies closer than 50 kpc. 
At $z<0.5$ there might be an excess of \prs with 3 galaxies within 50 kpc ($\sim10\%$). All these \prs have been found to live in X-ray selected COSMOS groups with  halo masses in the range $10^{13}-10^{14} M_{200c}/M_\odot$ \citep{george11}. 
Enlarging the radius of interest, the number of companions around \prs increases and a dependence on redshift might appear for the most extreme galaxies with the largest number of companions. Going from  higher to lower redshift, distributions shift toward a larger number of companions, suggesting that the environment around \prs gets richer. 
Considering our largest radius (500 kpc), we find that about 25\%  of \prs at 1.5$<$$z$$<$2 have no companions and another 25\% have at most two  companions. On the other hand, at lower redshift at most 5\% of \prs are isolated. At $z<0.9$, $\sim$45\% of  \prs with at least one companion have been found to live in X -ray selected COSMOS groups \citep{george11}. No group catalogs are available at higher redshift.

In Figure \ref{distr_sat} are also overplotted the same quantities for data drawn from the models. The upper panel shows the results for the \dl model, the lower panel for the H15 one.  At each redshift, we randomly extract the same number of \prs found in observations at the corresponding redshift, and then we compute the projected distances\footnote{For each galaxies we consider projected distances both on the {\it xy}, {\it xz} and {\it yz} planes. } to define the number of companions within a certain projected radius. We repeat the sampling ten times, to take into account sample variance. We then consider the maximum range spanned by the extractions. 
Qualitatively, within a large spread,  both the \dl and H15 models follow the observational trends, at all distances. Nonetheless, in both models there are extractions where the number of galaxies is systematically higher than observed, at all distances, especially in the \dl one. This might be related to the well known issue of the over-prediction of the number of satellites \citep[e.g.,][]{fontanot09, weinmann11} and has been found to be present also in clusters \citep{vulcani14}.  Discrepancies are solved by construction in the H15 model. 

The fact that in simulations there is a non-negligible spread among the different extractions indicates that the sample variance is not marginal, therefore a larger sample of observed galaxies will be needed to draw more robust conclusions. 

Our results are in line with the results of \cite{tal13}, who found that the total number of galaxies within a mass range of 1:10 and within roughly 400 kpc of the massive galaxy is on average 2-3 in all redshift bins \citep[see also][]{tal12, quilis12, gobat15}. % 

\subsection{Comparisons with the stellar mass function }\label{mf}
To further assess the environment in which the observed \prs live, we estimate how many satellites \prs should have based on the stellar mass function down to 1:10 the mass of the \pr. This number could give us hints whether most of them  live in clusters/groups or in the field. 

\cite{tal14} presented the stellar mass function of satellites around central galaxies of different stellar masses at $0.2 < z < 1.2$. They identified central galaxy candidates from the UltraVISTA catalog. Galaxies were considered to be central if no other, more massive, galaxies could be found within two projected virial radii. Virial radius estimates at a given stellar mass and redshift were determined using the semi-analytic model of \cite{guo11}.
They found that the mass distribution of satellite galaxies is independent of redshift for any given value of central galaxy mass. If we make the hypothesis that our \prs are the central galaxies of a group, we can integrate the \cite{tal14} mass function from the mass of the \pr down to 1:10 of its mass and compare the expected number to our observed ones. Analytically, we find that we should expect 2-3 companions per central galaxies within two projected virial radii.\footnote{Two virial radii is the size chosen by \cite{tal14}.} Following \cite{munoz11}, at $z\sim 1$, the typical virial radius of halos with $M_{vir}\sim 10^{13} M_\odot$ is $\sim 0.5$ Mpc. If we therefore consider a radius of 1 Mpc, we find that for $z>1$ 15\% of the \prs have at most one companion, while for $z<$1 there are no isolated \prs.
This might suggest that most of \prs live in massive structures like groups or clusters, but at least at high redshift there is a non-negligible fraction of them which have no companions.  

At $z<1$ we can push down the mass limit to 1:100, without being biased by mass-incompleteness. With this mass threshold, analytically, we should expect $\sim9$ companions per central galaxy within two projected virial radii. In our observations we find an average of 50$\pm5$ galaxies within the same radius. This might suggest that at least our low redshift \prs are located within structures largely dominated by  the presence of small galaxies. As we will see later on (\S\ref{sec:sfr+mer}), even though there is not much mass enclosed in these galaxies, they play a role in the mass growth of the \prs, given their high SSFR.

\subsection{Number of satellites as a function of distance}
\begin{figure}
\centering
\includegraphics[scale=0.4]{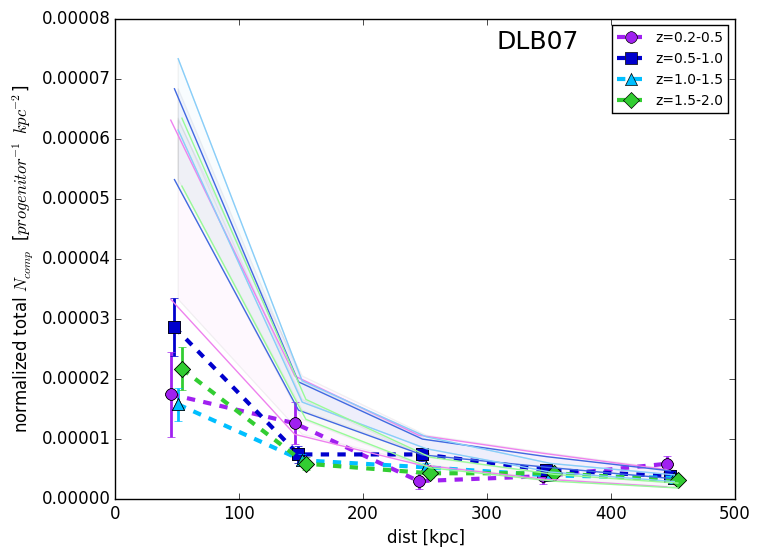}
\includegraphics[scale=0.4]{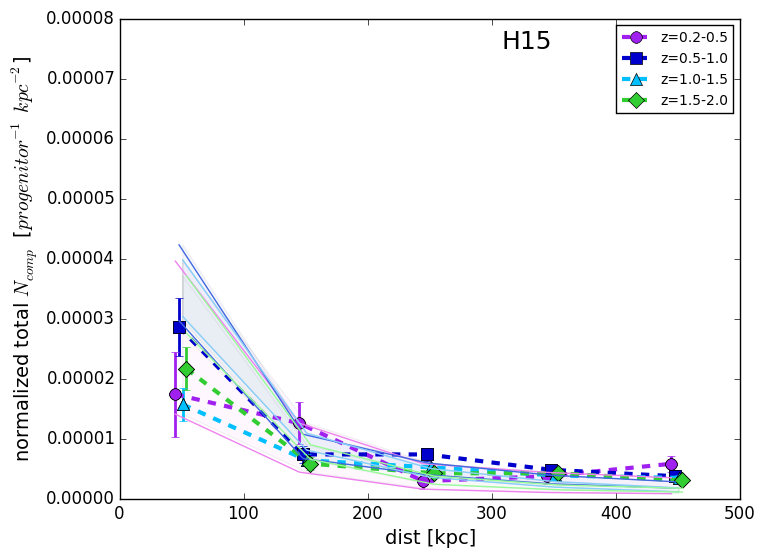}
\caption{Number of companions per \pr per kpc$^2$ as a function of distance at different redshifts, as indicated in the labels. Only \prs with at least one companion within 500 kpc, galaxies at least as massive as 1:10 the mass of the \pr and with z within $\pm0.05\times(1+z_{pr})$ are considered. 
Errors are poissonian. A horizontal shift is applied to the points for the sake of clarity.  Thick dashed lines and points represent observations; thin solid lines and shaded areas represent models (upper panel: DLB07, bottom panel: H15). In models, error bars represent the range spanned by the 10 extractions.  \label{sat_distance}}
\end{figure}

\begin{figure}
\centering
\includegraphics[scale=0.4]{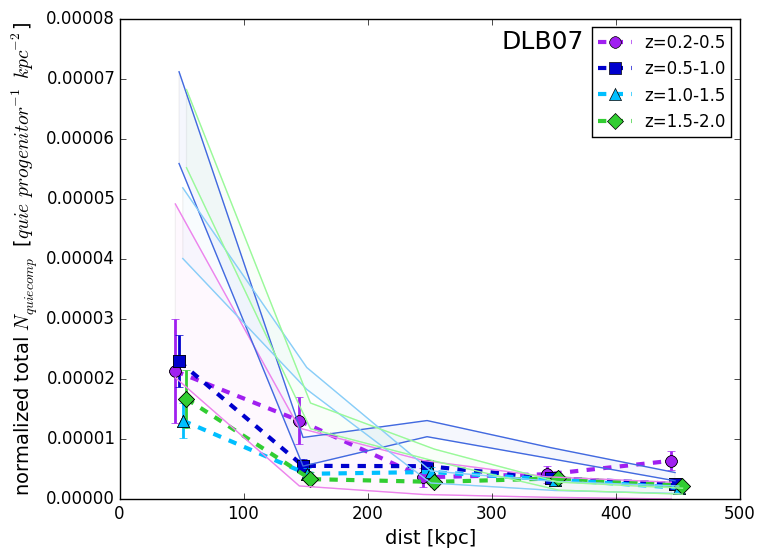}
\includegraphics[scale=0.4]{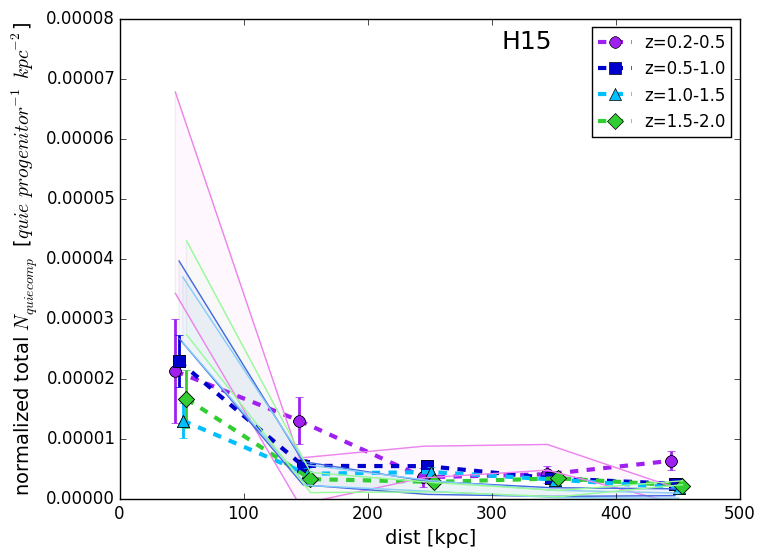}
\caption{Number of quiescent companions per quiescent \pr per  kpc$^2$ as a function of distance at different redshifts, as indicated in the labels. Points, lines, error bars and colors are as in Fig. \ref{sat_distance}
\label{sat_distance_type}}
\end{figure}

We now focus our attention on \prs with at least one companion within 500 kpc and investigate, on average, the variation of the number of companions per projected volume with distance, both in observations and in models (Fig.  \ref{sat_distance}). In observations, trends with redshift are not detected. 
However, if we sum up the number of companions, we find 4.6$\pm$0.9, 4.8$\pm$0.5, 3.7$\pm$0.3, 3.6$\pm$0.3  galaxies within 500 kpc per \pr from 0.2$<$$z$$<$0.5 to 1.5$<$$z$$<$2, respectively. 
Trends with distance are also detected.  The number of companions drops between distances of 100 and 200 kpc and is about constant at larger distances. 

Models show stronger trends with distance. In the \dl model (upper panel of Fig.\ref{sat_distance})  there is on average a larger number of companions around each \pr than in observations, especially at small distances (<100 kpc), where the number of galaxies  is more than a factor of two larger than in observations.
Considering 10 different random extractions, the median number of galaxies around \prs within 500 kpc  is 6$\pm$1, 7.4$\pm$0.5, 7.3$\pm$0.4, 5.9$\pm$0.3 from $z=0.36$ to $z=1.77$, respectively.  This again  might be due to the over-prediction of the number of satellites in the \dl prescription. In contrast, in H15 (bottom panel of Fig.\ref{sat_distance}),  the number of objects is typically consistent with the observed one (median numbers are 3.5$\pm$0.8, 3.7$\pm$0.4, 3.6$\pm$0.3, 3.1$\pm$0.2). In both models  no  trends with redshift are detected, even though the \dl model shows a possible inversion in the lowest redshift bin. 

Our findings are consistent with the idea that \prs are indeed centrals (so surrounded by a satellite population with some decreasing number density profile).  Indeed, as showed in Table \ref{tab_type}, in both models and at all redshifts, $>$90\% of the selected galaxies are classified as centrals in their halos. 

The discrepancies found in the \dl model  are noticeably alleviated if we exclude type 2 galaxies (plots not shown), which are mostly found at close distances from the center of the halos. Similarly, also in the H15 model the number of galaxies at $\sim$50 kpc is strongly reduced when type 2 galaxies are removed. 
We note, however, that type 2 galaxies cannot be excluded: it has been shown that their presence is fundamental to reproduce well several properties,  i.e. the clustering at small scales of the structures and the differences between the galaxy and subhalo profiles in the inner regions of  clusters \citep[e.g.][]{gao04, wang06}.

We can also investigate whether \prs with different star-forming properties live in different environments. Models do not provide us with the necessary information to distinguish between star-forming as quiescent galaxies as it was done for observations, nonetheless, both in models and observations we can use galaxy Specific Star Formation Rate ($SSFR=SFR/M_{\ast}$) to distinguish between star-forming and quiescent galaxies. We assume that galaxies with $\log SSFR<-11$ are quiescent, being this the minimum of the distribution. In observations, this number roughly corresponds to the adopted U-V vs. V-J cut.

In observations, as already mentioned, while at $z<1$ all \prs are quiescent, at higher redshift 60\% of \prs are star-forming.
Also in models the fraction of quiescent galaxies depends on redshift, spanning from $\sim 95 (78)\%$ to $\sim 30 (17)\%$ in the \dl (H15) model going from low to high-$z$.
Both in observations and simulations, there is no significant difference in the trends shown in Fig. \ref{sat_distance}  when we select only quiescent candidate progenitors.
 
Similarly, we now focus on the properties of the companions, checking whether their distribution around \prs depends on their star-forming  properties.
Figure \ref{sat_distance_type} shows the variation  with distance and redshift of the number of quiescent galaxies around quiescent \prs both in observations and simulations.  Trends resemble those found for the total population. In observations  the distance dependence is less steep, while in models there are hints it might be steeper.  
Both models show an over abundance of satellites at small distances. 
Being overall the trends similar in Fig. \ref{sat_distance} and \ref{sat_distance_type}, we can conclude that there are no evident signs of clustering in observations, where the quiescent and star forming galaxies are similarly distributed around \prs, while in models it seems that quiescent galaxies might be more clustered around quiescent progenitors. 
In observations, similar results have been obtained when using a U-V vs. V-J cut, showing that the results are not much sensitive to the cut adopted to separate star forming from quiescent galaxies.  

To summarize, we have found that the number of companions around \prs does not  depend on redshift. In observations, going from higher to lower redshift and from smaller to larger distances, the environment gets proportionally richer of galaxies. Nonetheless, there is a fraction of \prs that do not have companions, suggesting that their mass growth is hardly related  to merger events. Excluding isolated \prs, we found that the distribution of companions per projected volume is almost independent on redshift. 

In models, the fraction of isolated \prs is much lower, indicating \prs live in denser environments.

\section{The drivers of the progenitors' mass growth}
In this section we aim to investigate which are the most important factors that drive  the  galaxy mass growth for \prs from $z\sim 2$ to $z\sim 0$. We will first focus only on mergers (\S\ref{sec:mergers}), then we will quantify the contribution of  in-situ star formation (\S\ref{sec:sfr}) and finally combine the two (\S\ref{sec:sfr+mer}), to estimate their relative importance at the different redshifts.   

\subsection{What fraction of galaxy mass growth is due to mergers?}\label{sec:mergers}
\begin{figure*}
\centering
\includegraphics[scale=0.28]{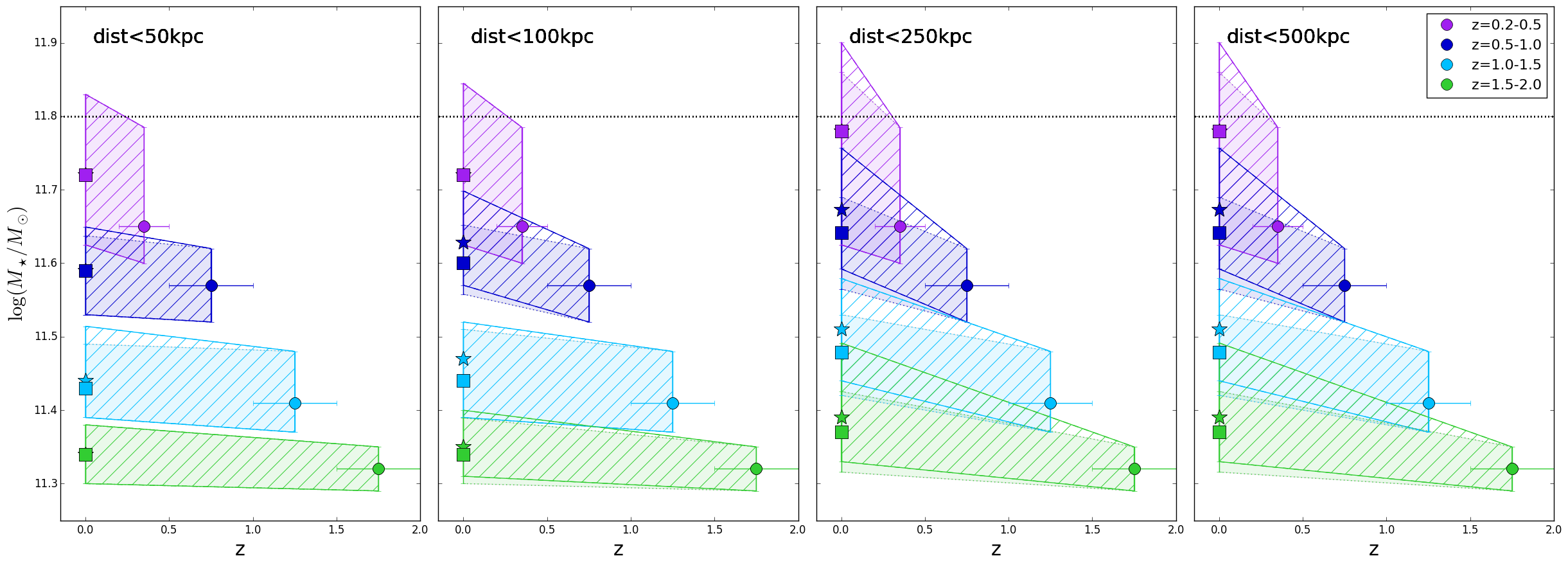}
\caption{Observed median \pr mass growth at different redshifts, as indicated in the labels, assuming that  galaxies within 50, 100, 250, 500 kpc will merge by $z=0$ onto the \pr. The mass at $z=0$ is  the sum of the entire mass (stars and dashed regions) or half of the mass (squares and shaded regions) of all galaxies with mass ratio 1:10 and redshift $\pm0.1$. Error bars on the x-axis represent the width of the redshift bin, error bars on the y-axis represent the 25$^{th}$ and 75$^{th}$ percentiles. The dotted horizontal lines represent the mass of the \umgs at $z\sim0$ \citep[from][]{marchesini14}.  \label{merge}}
\end{figure*}

\begin{figure*}[!h]
\centering
\includegraphics[scale=0.28]{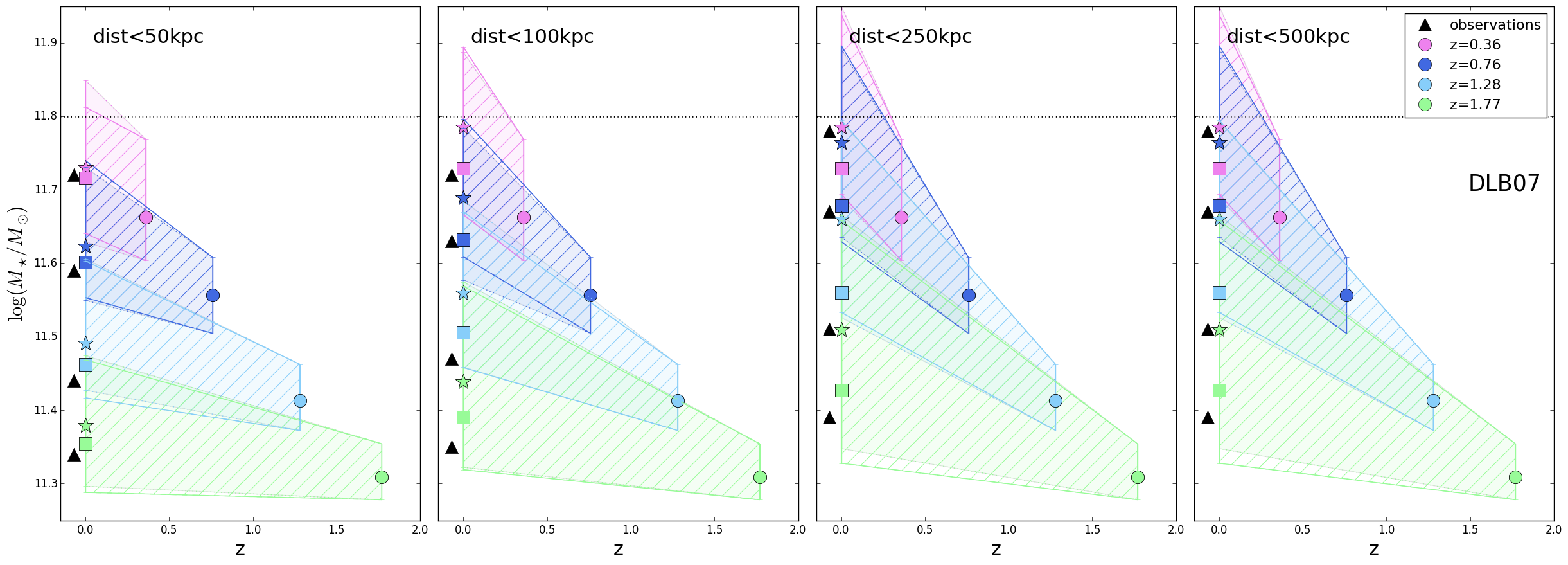}
\includegraphics[scale=0.28]{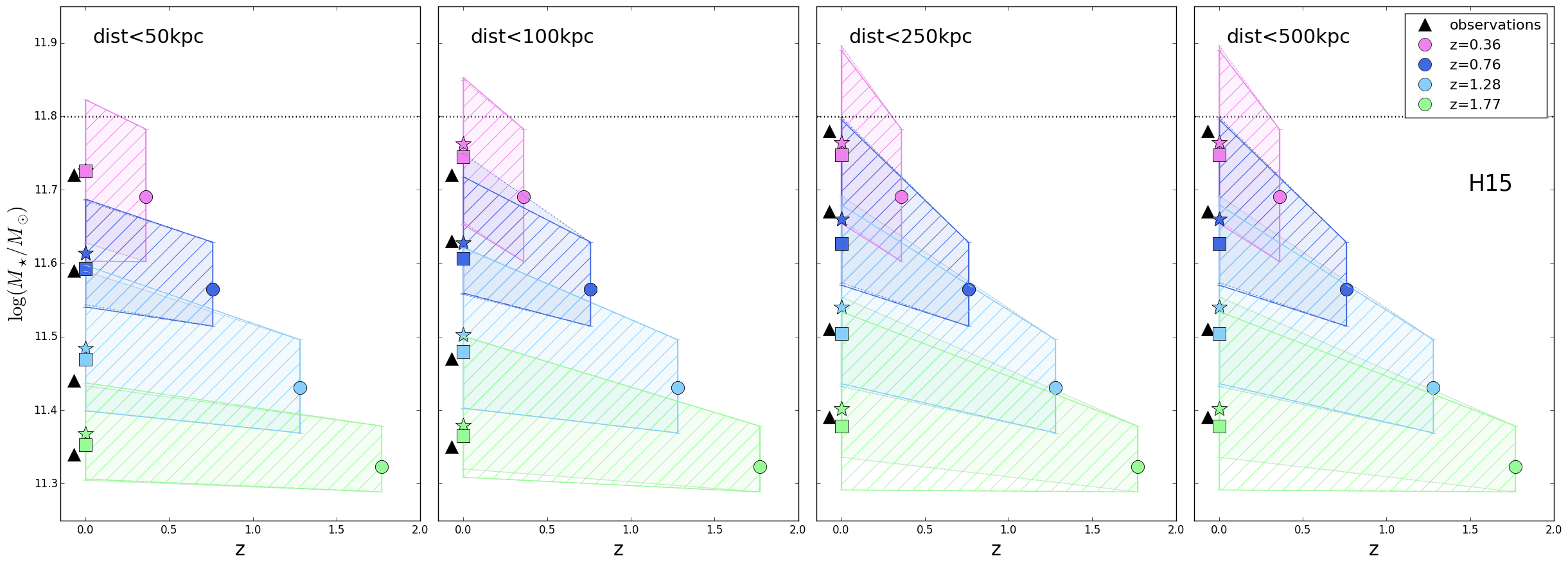}
\caption{Predicted median \pr mass growth at different redshifts, as indicated in the labels,  assuming that  galaxies within 50, 100, 250, 500 kpc will merge by $z=0$ onto the \umg. The mass at $z=0$ is  the sum of the entire mass (stars and dashed regions) or half of the mass (squares and shaded regions) of all galaxies with mass ratio 1:10. Error bars on the y-axis represent the maximum the 25$^{th}$ and 75$^{th}$ percentiles of the  10 representations. Upper panel: \dl, bottom panel: H15. The dotted horizontal lines represent the mass of the \umgs at $z\sim0$ \citep[from][]{marchesini14}. Black triangles represent the inferred mass at $z=0$ from observations, assuming that all the mass of the satellites fall into the \pr.  \label{sim_merge}}
\end{figure*}

\cite{kb08} investigated the major merger rates  using catalogs based on the \dl model to obtain the characteristic time-scale needed by two galaxies of a given mass ratio and redshift to  merge on the basis of their projected distance. In the model, to determine whether or not two galaxies might merge, it is assumed that when the subhalo which hosts a galaxy is tidally disrupted near the center of a more massive halo,  the galaxy becomes eligible to merge with the central galaxy of that halo. Nonetheless, the merger does not occur immediately, but rather after a ``dynamical friction time'' estimated from the relative orbit of the two objects at the moment of subhalo disruption. 

For $z\leq 1$,  stellar masses above $5\times 10^{9}M_\odot$ and  samples limited to radial velocity difference  $\Delta v<3000\,km/s$,\footnote{The authors suggest to use the timescales for $\Delta v<3000\, km/s$  when
analyzing data from photometric redshift samples, since the ``background''
correction will not eliminate physically associated galaxies at large velocity
separation.
}
\begin{equation}
  \label{eq:timescale}
  \langle T_{\rm merge}\rangle=3.2{\rm Gyr}\frac{r_p}{50{\rm kpc}}\ \left(\frac{M_*}{4.6\cdot10^{10}M_\odot}\right)^{-0.3}(1+\frac{z}{20})
\end{equation}
where $T_{\rm merge}$ is the time scale, $r_p$ the projected physical separation,  $M_*$  the  stellar mass of the pairs, $z$ the redshift of the \pr.

\cite{kb08} do not provide a formula for higher redshift galaxies, therefore we use the same parametrization also at 1$<z<$2. Given the range of masses and redshifts in our sample, we find that galaxies located more than $\sim$350 kpc apart at $z\sim2$ should not go through a merger event by $z=0$.

We can therefore estimate the mass growth due to mergers  by summing the stellar mass in companion galaxies that are expected to merge by $z=0$ with the mass of the \pr.{\footnote{For each pair \pr- companion we estimate the time they need to merge and evaluate if they will merge, eventually.} 
We analyze two cases: in the first one we assume that the entire mass in the companions will end up in the mass of the \pr, in the second case only half of it, while the other half is assumed to go into some diffuse component.
These are two extreme cases that should bracket the real situation \citep[e.g.,][]{conroy07,lidman13, burke15}.

Figure \ref{merge} shows the median mass growth that observed \prs are expected to have from their redshift to z$\sim0$.\footnote{We note that our values slightly differ from those presented in  \cite{marchesini14} because they adopt mean masses, whereas we use median values.} In the local Universe, the typical stellar mass of \umgs is $M_\star\sim 10^{11.8} M_\sun$ \citep{marchesini14}. 

We consider different bins of distance, and therefore only  galaxies within a certain distance from the \pr. 
Going to very large distances allows us to be as inclusive as possible in terms of companions to count and give an estimate of how much the mass growth changes as a function of distance. 

First of all, we note that the dispersion (described as the 25$^{th}$ and 75$^{th}$ percentile of the distributions) around the median mass is asymmetric, and is mainly due to the fact that we are on the exponential tail of the mass function, therefore distributions are not normal. 

The figure shows that the mass growth depends on the considered distance. Taking into account only galaxies within 50 kpc from the \pr, the mass growth is negligible for galaxies at $z>$0.5. In contrast, mergers alone might explain the growth of galaxies from $0.2<z<0.5$ to $z=0$. 
Increasing the radius of interest, the mass growth due to mergers increases; nonetheless it is generally  still insufficient to 
justify the expected mass growth. 
This is true both assuming that the entire mass in the companions will end up in the mass of the \pr, and that only half of it will. 

Taking into account all galaxies that can actually merge with the \pr from their redshift to $z=0$, on average, galaxies increase their mass of 41$\pm$9\%, 33$\pm$1\%, 27.9$\pm$0.4\%, 31.4$\pm$0.6\%  from z$\sim$ 0.35, 0.75, 1.25, 1.75 respectively.  

We note that only for $z<1$ we can include in the computation galaxies with smaller mass ratio (down to 1:100) with respect to the mass of the \pr, without being affected by sample incompleteness. Considering  also these galaxies in the computation does not strongly influence the results (plots not shown), simply because despite there are many more galaxies, their low-mass is negligible with respect the mass of the \pr.

When inspecting models (Fig. \ref{sim_merge}), similar results are obtained for the lowest redshift bin and the mass growth for galaxies at $z=0.36$ is compatible with the mass of the local \umgs.  In addition, the \dl model can explain the mass growth in terms of mergers from $z=0.76$ (1.28) to $z=0$ when a radius $\geq$100 (250) kpc is considered, the H15 one when a radius $\geq$250 kpc is considered.
We emphasize the large  spread that characterizes the models, which indicates the variety of growth histories that characterizes galaxies. 
%In general, both the \dl and the H15 models retrieve a stellar mass which is systematically higher than the one obtained from the observations, even though values are always compatible within the errors, given the large spread that characterizes the models. This is  a consequence of the larger number of companions already found, that might entail a larger number of mergers, under the hypothesis that all galaxies that are assumed to merge according to the \cite{kb08} formula do indeed merge. %, therefore, according to suggests that in models there is on average a than in observations. 

Recall that in the left panel of Fig. \ref{sim_growth}, where we compared the stellar mass of the \prs and their descendants as given by the models,  we found that in the H15 model the mass growth  is such that the descendants at $z=0$ do not have mass of $\sim 10^{11.8} M_\odot$, indicating that in this model the progenitor selection did not work properly. This is due to the fact that in the H15 model not all the mass in the merging galaxies ends up into the central one, but a not negligible fraction goes into the intracluster medium (H15). 

In summary, the analysis above suggests that  mergers alone can explain the mass growth of the \prs of the local \umgs only for galaxies at $z<0.5$ in observations, and  $z<1$, in simulations. In contrast,  they do not produce enough mass for galaxies at higher redshift. 
This is due to the fact that at $z>1$  \prs are isolated or have very low-mass companions, whose mass is not sufficient to explain the expected trends. 
These results suggest that other factors might play an important role in the galaxy mass growth. 

\subsection{What fraction of galaxy mass growth is due to star formation?}\label{sec:sfr}
\begin{figure*}
\centering
\includegraphics[scale=0.45]{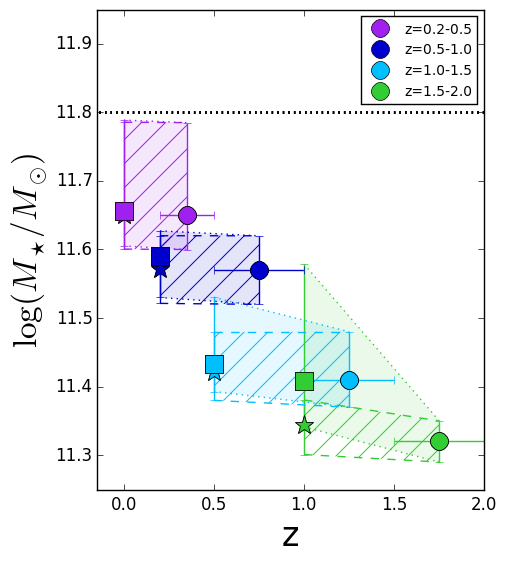}
\includegraphics[scale=0.45]{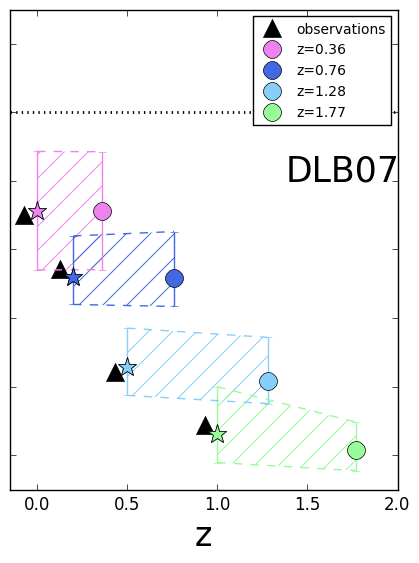}
\includegraphics[scale=0.45]{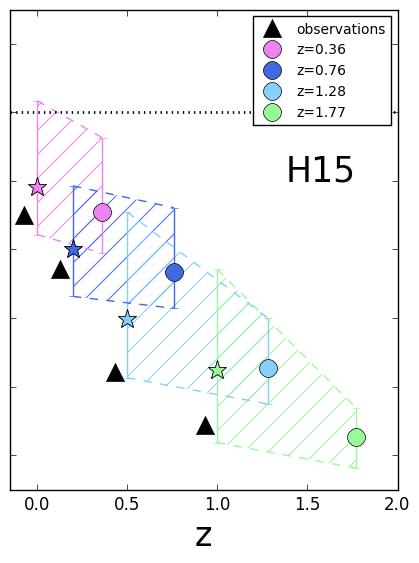}
\caption{Median \pr mass growth due to a constant star formation at different redshifts, as indicated in the labels. Left panel: observations. Stars and dashed regions show the mass growth adopting SFR estimates from SED fitting; squares and shaded regions show the mass growth adopting SFR estimates from UV+IR. Central panel: \dl model, right panel: H15 model. In the panels showing the models, black triangles represent the values obtained from observations.  Error bars on the x-axis represent the width of the redshift bin (only in observations), error bars on the y-axis represent  the 25$^{th}$ and 75$^{th}$ percentiles. 
 \label{sfr}}
\end{figure*}
Figure \ref{sfr} shows the amount of mass \prs are expected to gain for star formation, both in observations and models. In observations, we adopt both the SFR estimates obtained  from the SED fitting and those obtained from the UV+IR. The latter are systematically higher than those obtained from the SED fitting and give us an upper limit of the growth.  In simulations we use the SFR estimates provided by the two semi-analytic models.

We consider a constant SFR in the range of time between the redshift of the galaxy and the lowest limit of the next redshift bin. In this way our estimates most likely represent an upper limit of the real situation.\footnote{In observations, considering a declining SFR does not strongly change the results (plots not shown).} 
In the computation, we take into account the fact that the stellar mass of a galaxy changes with time also simply due to the evolution of its stars: as they progressively evolve and eventually die, they retain only part of their mass as remnant. Following \cite{poggianti13}, who used the \cite{BC03} model, the fraction of initial stellar mass that remains is equal to 1 for ages less than $1.9 \times 10^6 \, yr$, while it can be approximated as $f(t) = 1.749 - 0.124 \times \log t$ at older ages, where $t$ is the age of the stellar population in years. Approximately, in 0.6 Gyr galaxies retain $\sim$60-70\% of the mass they have formed.

The left panel of Fig. \ref{sfr} shows the results for observations. Estimates from SED fitting produce a little mass growth, while estimates from UV+IR can  explain the mass growth from one redshift bin to the next. 
In the \dl model (central panel of Fig. \ref{sfr}) the median values obtained are comparable to those observed from SFR estimated from the SED fitting. As in observations, star formation alone can only marginally explain the mass growth. In addition, results of the 10 different extractions are quite similar, as indicated by the moderately small scatter. This suggests that in the \dl model galaxies of similar mass
 have similar SFR at the time they have been selected.  In contrast, values in  the  H15 model (right panel) are systematically larger than the observed ones, even though there are some extractions where they show agreement. Note that in the H15 model the scatter is very large, indicating that galaxies of similar mass can have a wide range of SFRs.  

\subsection{Combining the contribution of SFR and mergers}\label{sec:sfr+mer}
\begin{figure*}
\centering
\includegraphics[scale=0.45]{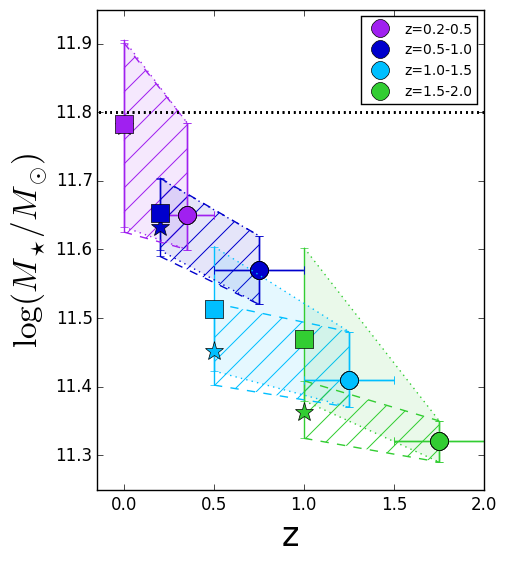}
\includegraphics[scale=0.45]{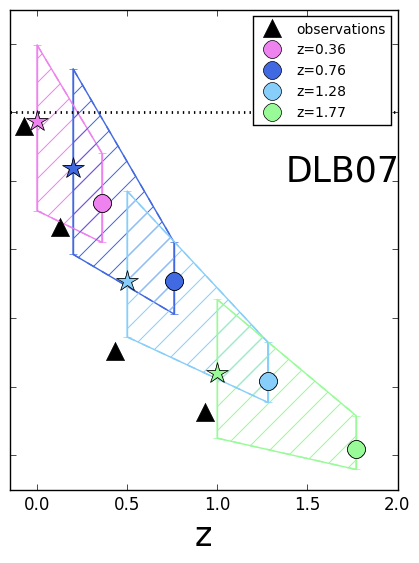}
\includegraphics[scale=0.45]{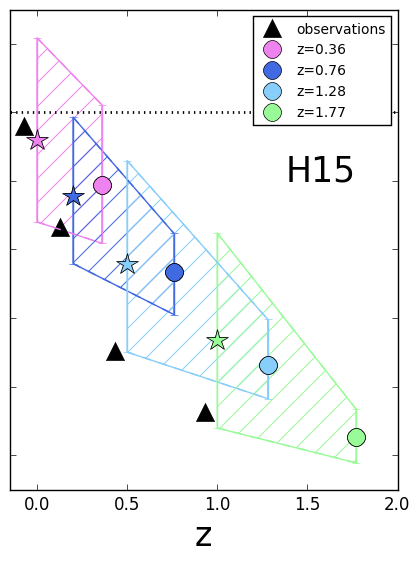}
\caption{Median \pr mass growth due to both star formation and merger at different redshifts, as indicated in the labels. Also the satellites' mass growth is taken into account. A constant SFR and the assumption that all the mass of the satellites will end up in into the mass of the \pr have been made. Left panel: observations. Stars and dashed regions show the mass growth adopting SFR estimates from SED fitting; squares and shaded regions show the mass growth adopting SFR estimates from UV+IR, central panel: \dl model, right panel: H15 model. In simulations,  black triangles represent the values obtained fro observations. Error bars on the x-axis represent the width of the redshift bin (only in observations), error bars on the y-axis represent  the 25$^{th}$ and 75$^{th}$ percentiles.  
 \label{sfr_mer}}
\end{figure*}

In the previous subsections we have found that neither mergers nor star formation alone are able to fully explain the expected \prs' mass growth from $z=2$ to $z=0$, both in observations and, to some extent, in simulations. Here we aim to test whether the combined contribution of in-situ and environmental processes can produce the expected growth. We also consider the mass growth due to star formation in the galaxies that will merge with the \prs.  We note that our analysis does not consider the contribution from starburst during mergers. This is probably not a dominant channel for mass growth, but might play a somewhat more important role at higher redshift.

As in the previous section, we compute the mass growth due to star formation in the time interval between the redshift of the galaxy and the next redshift bin. In this case we use the same time interval also to estimate the contribution of mergers, so that we can sum them up together. 
%In observations, {\bf for both \prs and satellites,} we use {\bf both} the SFR determined from the SED fitting {\bf and that obtained from the UV+IR}. %, which are available for all galaxies in the UltraVISTA sample. 
The left panel of Figure \ref{sfr_mer} shows the results for the observations. Considering all galaxies that might eventually merge,  when the SFR determined from the SED fitting is adopted, the combination of the two contributions 
marginally explains the mass growth, barely tracing the lower limit of the growth. Instead, when we adopt the SFR estimates obtained from the combination of UV and IR luminosities,  we recover the expected mass growth. Recall that the SFR$_{UV+IR}$ represents an upper limit of the true values, given that at these redshifts the AGN contamination might not be negligible. However, the SFR estimated from the UV+IR is arguably less biased against heavily obscured star formation. Therefore, the real growth is expected to be bracketed between these two cases.

The central and right panels of the same Figure show the results for the two semi-analytic models. Both predict a growth sufficient to support the expected mass growth, at all redshifts. Again, the large scatter that characterizes the models suggests that the different extractions we performed from the catalogs can give quite different results.

The result obtained in the right panel for the H15 model is in disagreement with the results presented for the same model in Fig.\ref{sim_growth}. 
When we apply the same prescriptions as done in the observations to account for the growth from both merging and in-situ star formation from one redshift bin to the following one, the inferred mass  is in agreement with the mass of the \prs at that redshift. As a consequence, we can grow all progenitors to a mass of $\sim 10^{11.8} M_\odot$ at $z=0$. In contrast, when we directly consider the mass of the descendants at $z=0$ as provided by the models, we find a systematically smaller mass for the galaxies of the same initial mass. This means that some of the assumptions  made to estimate these contributions may not be sufficient or may be incorrect (e.g., stripping and/or merger rates) when adopted for this particular model, as it will be discussed in the following section.  

\section{Discussion}
\begin{figure*}
\centering
\includegraphics[scale=0.32]{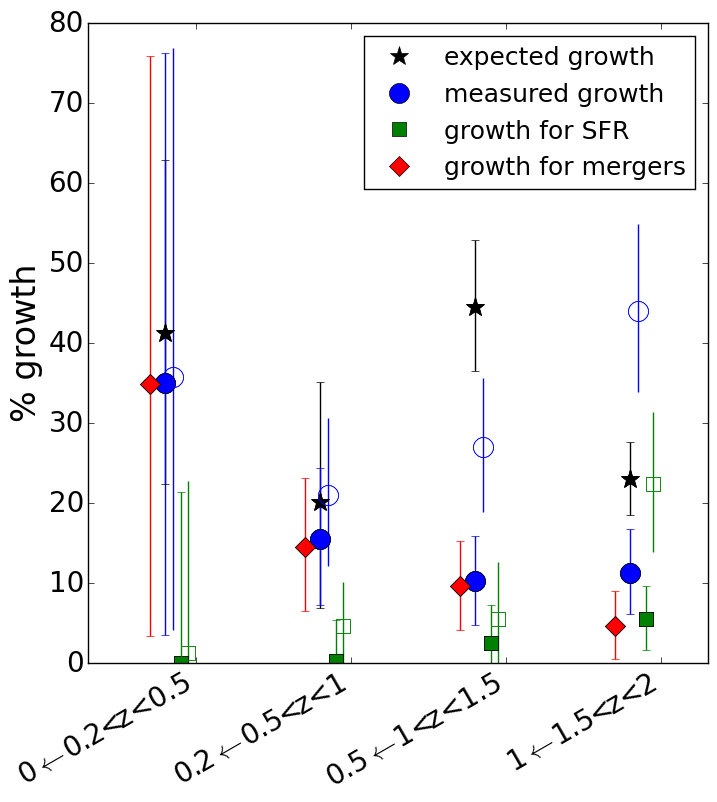}
\includegraphics[scale=0.32]{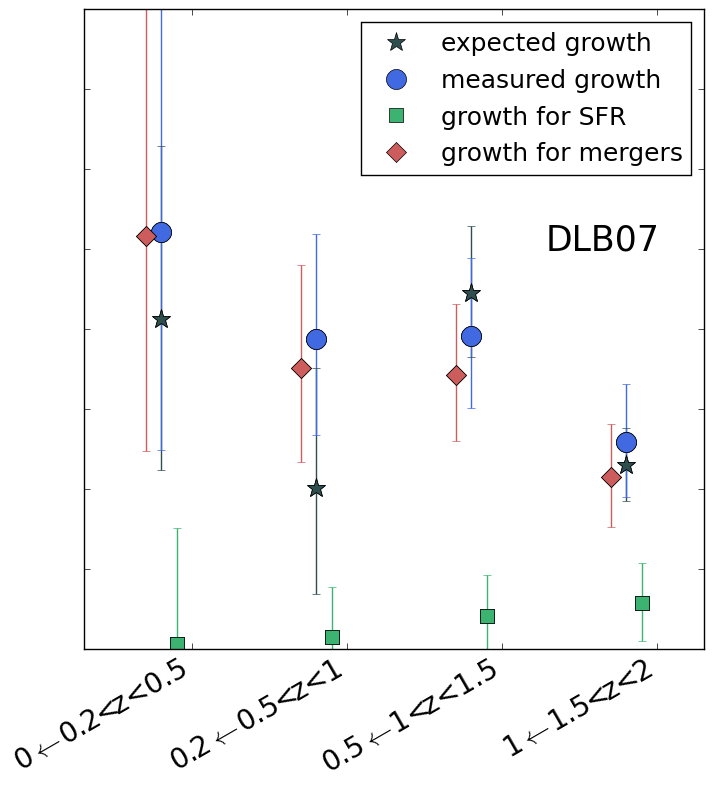}
\includegraphics[scale=0.32]{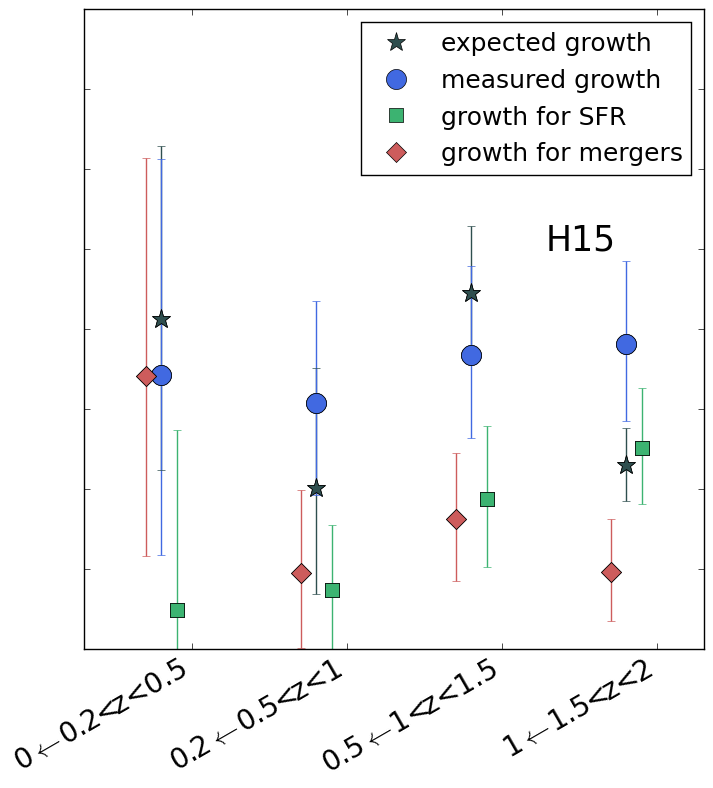}
\caption{Percentage of the mass growth as a function of time as expected by the abundance matching technique (black stars), as measured considering the combined contribution of SFR, mergers and SFR in the satellites that will merge (blue points), as measured considering only SFR in the \prs (green squares) and only mergers (red diamonds), for observations (left panel), the \dl model (central panel) and the H15 model (right panel). In observations, values obtained both considering the SFRs from the SED fitting (filled circles) and those from the UV+IR (empty circles). A shift has been applied to the points for the sake of clearness. Error bars represent the maximum and minimum growth, obtained propagating the errors on the medians \citep{rider60}. 
\label{summary}}
\end{figure*}

\begin{figure*}
\centering
\includegraphics[scale=0.35]{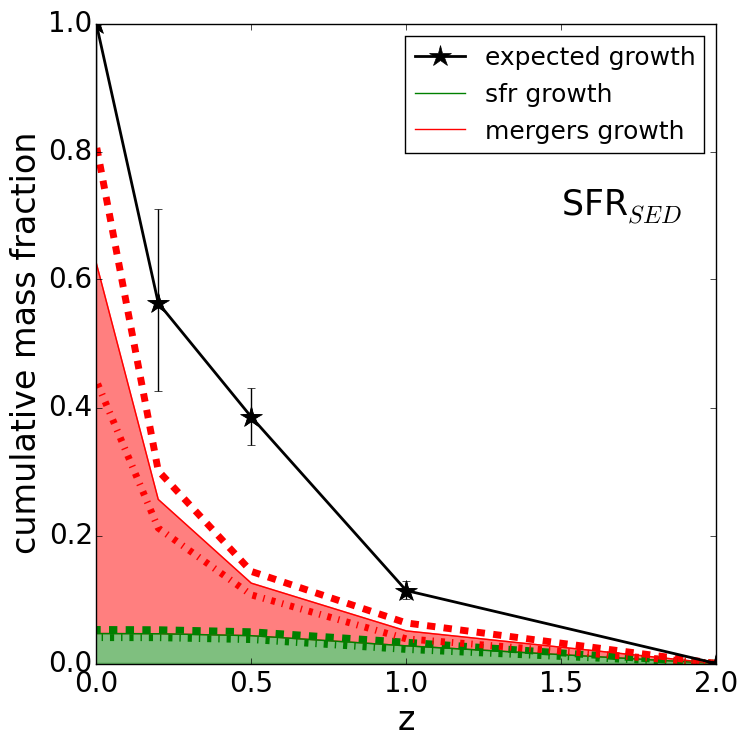}
\includegraphics[scale=0.35]{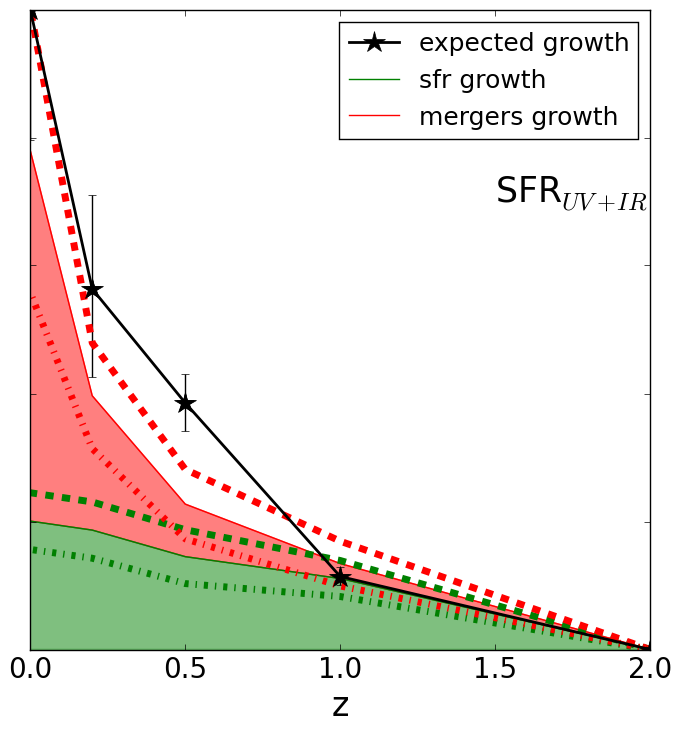}
\includegraphics[scale=0.35]{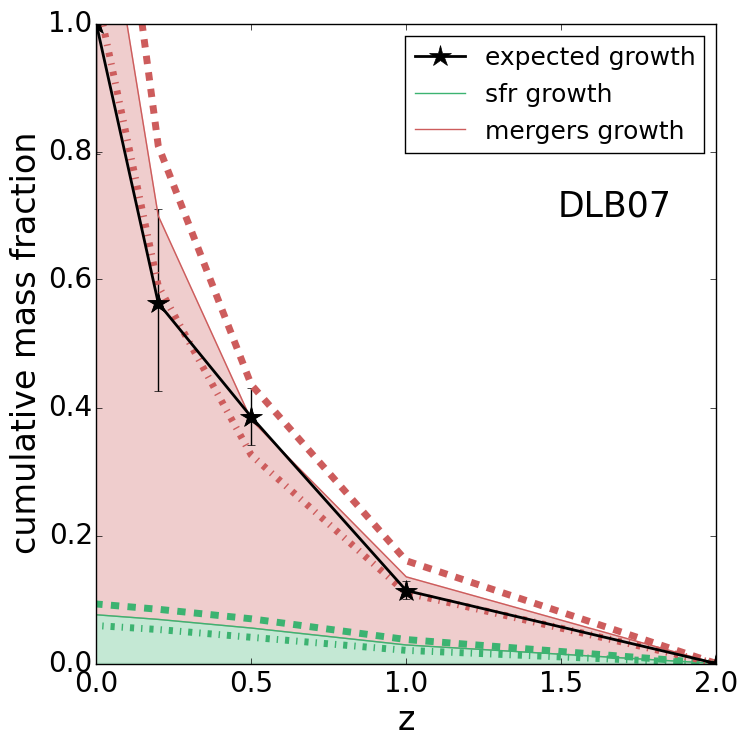}
\includegraphics[scale=0.35]{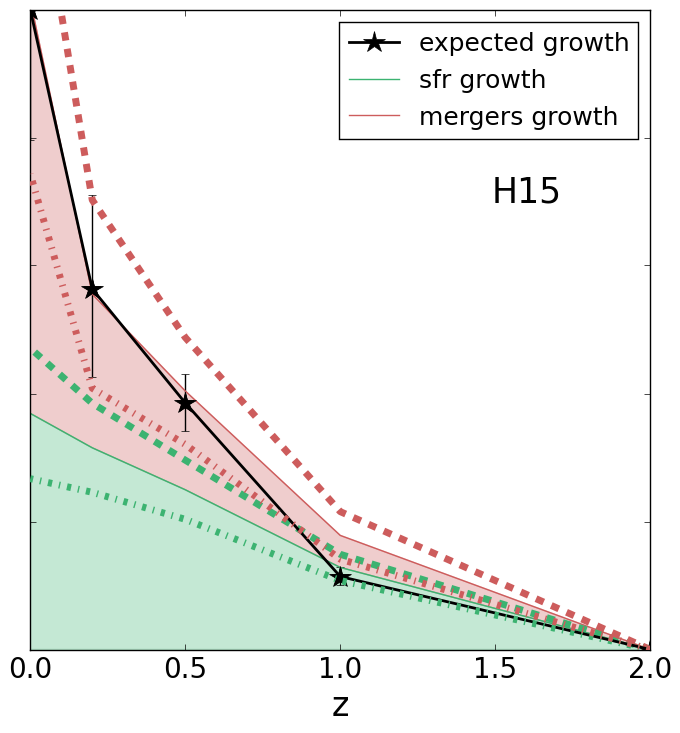}
\caption{Mass build- up over time due to star formation and mergers. Upper panels: observations;  left panel: SFR from the SED fitting, right panel: SFR from the UV+IR luminosities.  Bottom panels: simulations; left panel: \dl, right panel: H15. Black stars and solid lines represent the expected mass growth from \cite{marchesini14} with the uncertainties; red regions the contribution of mergers to the mass growth, green regions the contribution of SFR to the mass growth. Thick dashed lines represent the upper limit of the contributions, while thick dash-dotted line the lower limit, both obtained  propagating the errors on the medians \citep{rider60}.
\label{summary2}}
\end{figure*}

The main results of our analysis are summarized in 
Figures \ref{summary} and \ref{summary2}. The former presents the relative contribution of star formation and mergers to the mass growth, showing the inverse of ratio of the \pr stellar mass at a given redshift to the inferred stellar mass at the same redshift. The expected stellar mass growth from the abundance matching technique, the total measured stellar mass growth obtained considering mergers and star formation and the separate contribution of star formation and mergers are shown. The latter figure shows the cumulative mass growth due to the two contributions, separately. In both figures, the quoted uncertainties represent the maximum and minimum growth, while  errors on the medians are estimated as $1.253\sigma/\sqrt{N}$, where $\sigma$ is the standard deviation about the median and N is the number of galaxies \citep{rider60}. 

%First of all, we note that i
In observations (left panels) the total mass growth we obtained is below the expectations, when the SFR values are obtained from the SED fitting}. Trends are driven by high-$z$ galaxies: from $1<z<1.5$ to $z\sim0.5$ we measure a growth of $\sim 10\%$ while the expected growth is of $\sim 40\%$. This entails that at $z=0$ $\sim 20\%$ of mass is lacking.  In contrast, discrepancies are largely reduced when SFRs are measured from a combination of UV and IR luminosities. In this case, the inferred mass growth is even larger than the expected one at the highest redshift. 

Focusing on models (central and right panels), both  prescriptions are able to fully explain the mass growth as predicted by the abundance matching technique, and, possibly,
even over-predicting it.

Investigating separately the contribution of
star formation and mergers, we find that they play a very different role at the different redshifts. 
In observations, the average star formation rate is similar to the net growth rate at $z = 1.5-2$ but significantly smaller at later times. 
Progenitor's star formation is only important at the highest redshifts, where it might be able to explain alone all the mass growth. 
At lower redshifts mergers acquire importance, and they are the major responsible of the observed evolution at $z<0.5$. At the intermediate redshifts, the growth can be explained advocating the combined contribution of star formation and mergers.  

Overall, these findings are in line with many other studies \citep[e.g.,][]{vd99, vd05, vdokkum10, tran05, bell06, white07, mcintosh08, naab07, naab09,  
ownsworth14}, even though some works  have suggested that major mergers may play a more prominent role with up to $\sim 60\%$ of a massive galaxies stellar mass growth at $z < 2$ arising from major merger events \citep[e.g.,][]{lopez12, ferreras14, ruiz14}.

In both models the contribution of  star formation to the total mass growth decreases with time. However, it plays a larger role in the H15 model  than in the \dl, at all redshifts,  and it shows a steeper decline with time. It goes from  20\% to 5\% in the H15 model, from $<10\%$ to 0\% in the \dl model. 
In the \dl model the contribution of mergers increases with time, ranging from 20\% to 50\%, with a slope that is similar to the observed one. However, differently from observations, they are much more important than star formation even at higher redshift.
In contrast, in the H15 model the contribution of mergers is roughly constant with time, showing a bump only in the lowest redshift bin. Values are similar to the observed ones.

As already mentioned, the non-negligible spread measured  among the different extractions in simulations indicates that the sample variance is not marginal, therefore a larger sample of observed galaxies will be needed to draw more robust conclusions. 

In models, we can explicitly investigate the separate role of mergers and star formation by inspecting the merger trees of a subsample of galaxies in the highest redshift bin. We consider only \prs at $z=1.77$, since most of the \prs at lower redshift are actually descendants of these galaxies, therefore will enter their merger trees later on.
\begin{figure*}
\centering
\includegraphics[scale=0.35]{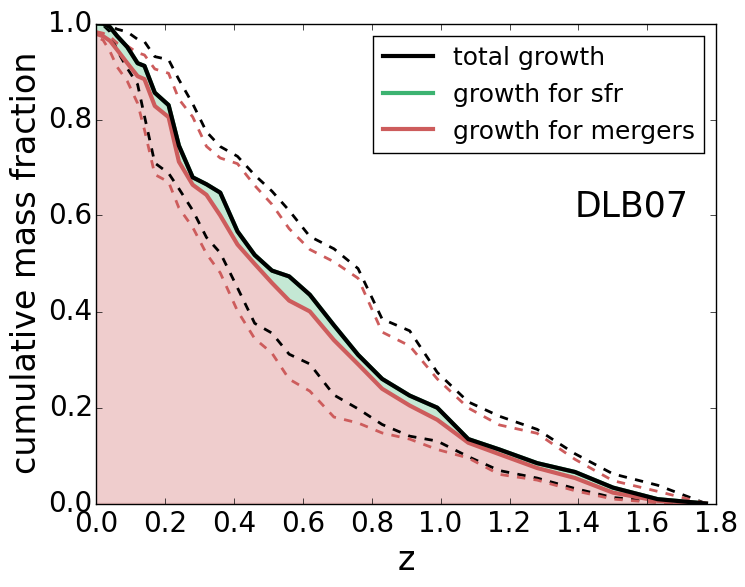}
\includegraphics[scale=0.35]{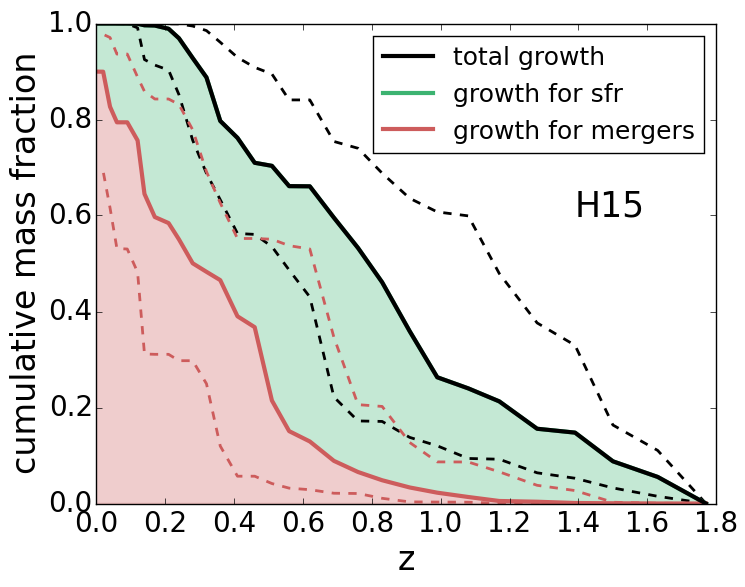}
\includegraphics[scale=0.35]{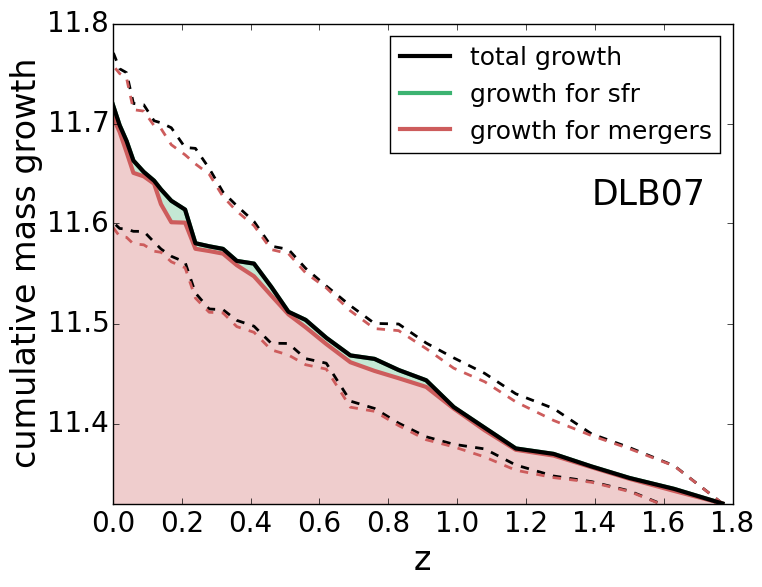}
\includegraphics[scale=0.35]{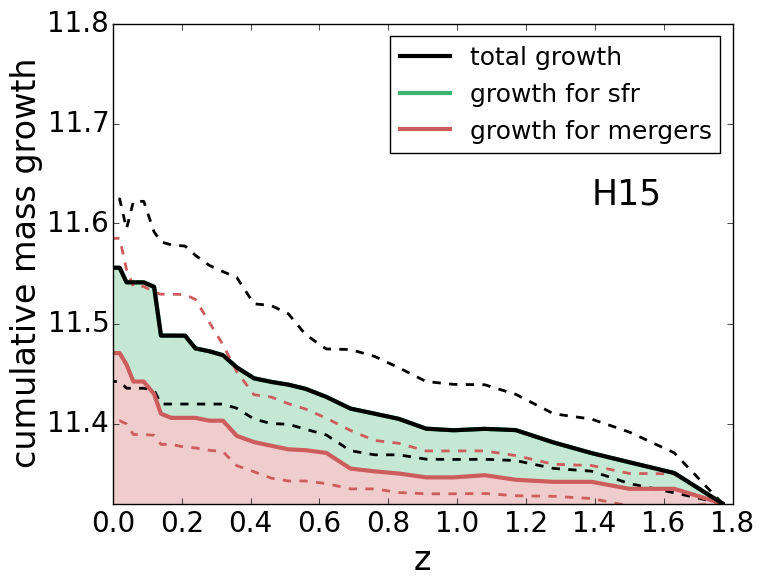}
\caption{Predicted mass build-up over time due to star formation and mergers, as obtained from the merger trees, for the DLB (left panels) and H15 (right panels) models. Upper panels: cumulative mass fraction, lower panels: cumulative mass growth. The median of 70 merger trees is shown (solid lines), along with 1$\sigma$ dispersion (dashed lines). Black lines: total growth, red lines and areas: growth due to mergers, green lines and areas: growth due to star formation. See text for details. 
 \label{mt}}
\end{figure*}

In both models, for each \pr  we select all the galaxies in the merger tree that have its same direct descendant in the next snapshot and we sum up the masses of the merging galaxies to compute the mass growth due to mergers. We then compare the mass of the merged galaxies to the mass of their unique descendant  and assume that the difference is due to star formation.\footnote{Note that this means the stars formed during star-bursts associated with mergers are going to be in the mass growth phase}
Note that in the H15 model we consider only the galaxies that indeed merged, and not those disrupted before merging onto its descendant, whose matter went into the intracluster medium.
We then start from the descendant and repeat the loop down to $z=0$, in order to trace the 
 entire growth.

Figure \ref{mt} shows the median relative mass growth and the median total mass growth for 70 galaxies\footnote{This is the number of the observed progenitors at 1.5$<z<$2.} in the \dl and H15 models, respectively. As found in the previous sections, the contribution of mergers and star formation is different in the two prescriptions, being mergers relatively more important in the \dl model than in the H15 one. In H15 star formation plays an important role also at low redshift. Most importantly, we find that in the two models  galaxies are characterized by an overall very different mass growth. In line with the results shown in Fig. \ref{sim_growth}, from $z=1.77$ to $z=0$ galaxies in the \dl model growth $\times1.5$ more than in the H15 model. 
This is likely due to the modifications in the H15 model, which add the tidal stripping and reduce the mass of the merging satellites, therefore producing a smaller mass growth. Nonetheless, this result is quite surprising given the fact that the model reproduces the evolution of the stellar mass function and  has been calibrated to be in better agreement with the halo occupation distribution results (H15), which the abundance matching technique relies on. Understanding the reasons of these discrepancies is beyond the scope of this work, and is deferred to a forthcoming work. 

\subsection{Some caveats}
Overall, our analysis reveals some tensions between the mass growth expected by the abundance matching technique and the mass growth measured taking into account in-situ star formation and mergers, both in simulations and in observations. 
 The largest discrepancy is seen between $1<z<1.5$ and $0.5<z<1$ in observations, when the SFRs from the 
SED fitting are used. 
We note that these two contiguous redshift bins bracket a break in the galaxy property distribution: at $z>1$ many progenitors and satellites are still star forming, while at $z<1$ all the progenitors and most of the satellites are quiescent \citep[see also][]{marchesini14}.  As we will see later on, this transition regime might be responsible for this gap.

Overall,  discrepancies might be due to a number of factors. First of all, it might be that the  abundance matching technique adopted  to link galaxies across time does not work properly. Anyway, 
our analysis shows that candidates are selected well, at least for \dl model. Quite surprisingly, the H15 does not support the selection via the abundance matching technique. 
Indeed, the mass growth estimated for the H15 model and the one that is intrinsic in the model (as shown in Figures 13 and 2) are not in agreement. 
Understanding the weakness of the selection criteria is beyond the scope of this paper, since the method has been largely discussed in the literature and it has been found to provide an excellent match to a number of galaxy clustering statistics at multiple epochs \citep{kravtsov04, tasitsiomi04, vale04, vale06, conroy06, berrier06, marin08, trujillog11} and to a number of population properties \citep[e.g.,][]{conroy09, drory08}.

Alternatively, discrepancies might be due to  some assumptions made. For example, 
it might be that in reality mergers play a less important role at higher redshift than that estimated by \cite{kb08} at $z<1$. Even though merger rates are not expected to vary much with redshift \citep[e.g.,][]{guo08, kb08, wetzel09}, we might be 
over-estimating the number of mergers. %However, testing the adopted formula is beyond the scope of this paper, and it will be investigated in a future work.

In addition, in our treatment, we are not considering some other factors that indeed might play a role. 
The most important is the contribution of galaxies whose mass is lower that 1:10 the mass of the \pr and therefore do not enter our selection. These galaxies can be characterized by high SSFR values, hence they double their mass rapidly, therefore giving a non-negligible contribution to the total growth. We checked that at least at $z<1$, where our sample is not affected by incompleteness, including in the computation all galaxies with a mass ratio of 1:100 better reconcile the expected to the observed growth (plot not shown). We can not extend to higher redshift because of incompleteness effects.  We note, however, that at $z>1$ the contribution of satellites with mass ratio larger than 1:10 might play a more important role than that  at lower redshift. Indeed, most of them are star forming and are probably  characterized by high SSFR values, therefore giving a large contribution to the total growth. At $z<1$ many satellites are quiescent and contribute less to the total growth. 

Additionally, mergers can also induce bursty events of star formation, which can pump up  galaxy masses. However, it is very hard to  properly model these bursts  and quantify their role in the overall galaxy growth. 
  Not considering the contribution from starburst during mergers has a larger impact at $z>1$ that at $z<1$. Indeed, at higher redshift mergers most likely involve star forming galaxies and are accompanied by bursts that enhance galaxy star formation, with a consequent larger mass growth, while at lower redshift, mergers most likely take place between quiescent galaxies, therefore bursts are rare.

Finally, in observations, uncertainties in the star formation histories, dust content and distribution, the IMF, and other effects can easily introduce systematic errors of a factor of $\sim$2 in the star formation rates, particularly at high redshift \citep[see, e.g.,][]{reddy08, wuyts09, muzzin09}. Even though nowadays there is reasonable agreement between the global stellar mass density inferred at any particular time and the time integral of all the preceding instantaneous star-formation activity, modest offsets may still point toward systematic uncertainties  that are not negligible \citep[see][for a review]{madau14}.

Moreover, the tensions seen between $1<z<1.5$ and $0.5<z<1$ in observations are at least partly due to the fact that our analysis relies on COSMOS data, which  covers only one field of view. Therefore, we are not able to control for sample variance. \cite{guzzo07} identified a large-scale structure at $z\sim$0.73, which certainly contaminates the counts at $0.5<z<1$, most probably having an impact on the mass functions and on the cumulative number densities  involved in the selection of the progenitors. Having a larger sample of galaxies, based on several fields, is mandatory to really prove the existence of the observed gap and to understand its origin. 

\section{Summary and Conclusions}
The aim of this paper was to 
test the model predictions for the different contributions to the stellar mass assembly since $z\sim$2, and investigate the role of the star formation and mergers at the different redshifts. 
We  compared observational results with the data of two different semi-analytic models, to obtain a better insight on the physical processes responsible for the evolution. 

First, we characterized the environment of the \prs of local \umgs at $0.2<z<2$, selected with a semi-empirical approach using abundance matching in the $\Lambda$CDM paradigm \citep{behroozi13, marchesini14}. 
We investigated the number of companions around each \pr, in order to give an estimate of the environment surrounding these massive galaxies. The number of galaxies with mass at least 1:10 the mass of the \pr and with redshift within $\pm0.05\times(1+z_{pr})$ around \prs depends on distance. In observations, at any redshift $\sim80\%$ of the \prs have no galaxies within a projected radius of 50 kpc. This number drops to 25\% at $z\sim1.75$ and 5\% at lower redshift when a radius of 500 kpc is considered. In general, going from higher to lower redshift the environment gets proportionally richer of companions. 
Models qualitatively agree with observations, even though the fraction of isolated \prs is much lower in the models, indicating \prs live in denser environments, pointing to the well known over-estimation of satellites at high redshift (in the \dl model). 

Considering only \prs with at least one companion within 500 kpc, in both observations and simulations the number of companions decreases with distance. Nonetheless, the \dl model overestimates the number of companions almost at all distances \citep[in agreement with previous results, e.g.][]{weinmann11, vulcani14}. The star-forming properties of \prs and companions seem not to influence the trends in observations, while in models the fraction of quiescent companions might be higher around quiescent \prs. 

In the second part of the paper we investigated which are the most important factors in the \prs' mass growth at the different redshifts, characterizing the separate contribution of star formation and mergers. 

Overall, our analysis confirms the model predictions, showing how the growth history of massive galaxies is dominated by in situ star formation at $z\sim2$, both star-formation and mergers at $1<z<2$, and by mergers alone at $z<1$.
Nonetheless, detailed comparisons reveal some tension between the mass growth expected by the abundance matching technique and that measured both in observations and in simulations. 
In observations  we  recover a systematically smaller mass growth  when SFRs from SED fitting are adopted, whereas we obtain an overall comparable mass growth when SFRs from UV+IR are used. The true mass growth has to be bracketed between these two cases. %especially from $1<z<1.5$ to $z\sim0.5$, where we measure a growth of $\sim 10\%$ while the expected growth is of $\sim 40\%$. 
In models, both the prescriptions explain the mass growth as predicted by the abundance matching technique, and, when errors are taken into account even over-predict it. The role of the different contributions is different in the two prescriptions, highlighting how much the mass growth is model dependent. It is worth noticing that the mass growth estimated for the H15 model and the one that is intrinsic in the model are not in agreement. This implies that at least some of the assumptions made to estimate the different
contribution to mass growth might be wrong.  

Discrepancies might be due to a number of factors, such as an incorrect progenitors-descendants selection, an underestimate of minor mergers ($>1:10$), the adopted assumptions on  merger rates, or uncertainties on star formation rate and mass estimates.

In the future, a larger sample of observed galaxies will be needed to draw more robust conclusions.  Indeed, the non-negligible spread measured  among the different extractions in simulations indicates that the sample variance is not marginal. A larger sample would also allow to better investigate the impact of the environmental processes on galaxy evolution. It has been shown that the efficiency and the time scale of quenching of star formation in satellites is halo mass dependent, therefore at any redshift the role of star formation is certainly different in different environments \citep[e.g.,][]{dekel06, dekel09}. 

Another natural step forward to this analysis will be to better characterize 
the gas content and the interstellar medium (ISM) in the progenitors' population, especially in the star-forming progenitors, during a period in cosmic history that is most critical for the formation of their stars ($1<z<3$). Observation e.g. with ALMA will allow to carefully investigate  the ISM, which is the crucial ingredient fueling the activities of star formation and AGNs and separating the two contribution will better constrain the actual role of star formation in the total mass growth.

\section*{Acknowledgments}
We thank the referee for her/his comments that helped us to improve the manuscript. 
B.V. acknowledges the support from the World Premier International Research Center Initiative (WPI), MEXT, Japan and  the Kakenhi Grant-in-Aid for Young Scientists (B)(26870140) from the Japan Society for the Promotion of Science (JSPS).
D.M. acknowledges the support of the Research Corporation for Science Advancement's Cottrell Scholarship.
B.M.J. acknowledges support from the ERC-StG grant EGGS-278202.
 The Dark Cosmology Centre is funded by the DNRF.
This study is based on data products from observations made with ESO Telescopes at the La Silla Paranal Observatory under ESO programme ID 179.A-2005 and on data products produced by TERAPIX and the Cambridge Astronomy Survey Unit on behalf of the UltraVISTA consortium.
The Millennium Simulation databases used in this paper and the web application providing online access to them were constructed as part of the activities of the German Astrophysical Virtual Observatory (GAVO).

\bibliographystyle{apj}
\bibliography{biblio_UMG}

\label{lastpage}
\end{document}